\documentclass[twocolumn,aps,prb]{revtex4-1}

\pdfoutput=1

\usepackage{amsmath,amssymb}
\usepackage{graphicx,epsfig,sidecap,wasysym}
\usepackage{color}
\usepackage[colorlinks=true, urlcolor=blue, linkcolor=red, citecolor=blue, pdftex]{hyperref}
\hypersetup{pdfstartview={FitV}, pdfpagemode = UseNone, pdfpagelayout=SinglePage, bookmarksnumbered={true}}

\newcommand{\fullket}[1]{\left|#1\right>}
\newcommand{\fullbra}[1]{\left<#1\right|}
\newcommand{\twopartdef}[4]{\left\{\begin{array}{lll}#1 & \mbox{if } #2 \\#3 & #4\end{array}\right.}
\newcommand{\psistate}{\Psi}

\newcommand{\calH}{{\cal H}}
\newcommand{\calL}{{\cal L}}

\newcommand{\calC}{{\cal C}}

\newcommand{\Neel}{{N\'{e}el}}
\newcommand{\const}{{\rm const.}}

\begin{document}

\title[Valence bond distribution and correlation
in bipartite Heisenberg antiferromagnets]{Valence bond distribution and
correlation in bipartite Heisenberg antiferromagnets}

\author{David Schwandt}

\author{Fabien Alet} 

\affiliation{Laboratoire de Physique Th\'eorique,
IRSAMC, Universit\'e de Toulouse, CNRS,
31062 Toulouse, France}

\author{Masaki Oshikawa}

\affiliation{Institute for Solid State Physics, University of Tokyo,
Kashiwa 277-8581, Japan}

\date{\today}

\begin{abstract}
Every singlet state of a quantum spin-$1/2$ system can be decomposed
into a linear combination of valence bond basis states.
The range of valence bonds
within this linear combination as well as the correlations between them
can reveal the nature of the singlet state, and are key ingredients in
variational calculations. In this work, we study the bipartite valence bond
distributions and their correlations within the ground state of the Heisenberg
antiferromagnet on bipartite lattices.
In terms of field theory, this problem can be mapped to
correlation functions near a boundary.
In dimension $d\geq 2$, a non-linear $\sigma$ model analysis reveals
that at long distances the
probability distribution $P({\bf r})$ of valence bond lengths decays
as $|r|^{-d-1}$ and that valence bonds are uncorrelated.
By a bosonization analysis, we also obtain $P({\bf r})\propto |r|^{-d-1}$
in $d=1$ despite the different mechanism.
On the other hand, we find that correlations between valence bonds
are important even at large distances in $d=1$, in stark contrast
to $d \geq 2$. The analytical results
are confirmed by high-precision quantum Monte Carlo simulations in
$d=1$, $2$ and $3$. We develop a single-projection loop variant of the
valence bond projection algorithm, which is well designed to compute
valence bond probabilities and for which we provide algorithmic
details.
\end{abstract}

\maketitle

\section{Introduction}

Magnetic ordering is a primary concern of classical
magnetism. Long-range order can be disrupted by the presence of
disorder or of frustration between the magnetic degrees of
freedom. Quantum fluctuations form also an important path to destroy
long-range order, specifically antiferromagnetic (AF) order. They are
particularly strong for low values of the spin $S$ and in low
dimensionality. The prototypical model to study quantum fluctuations is
the $S=1/2$ AF Heisenberg model
\begin{equation}\label{eq:H}
{\cal H} = \sum_{\langle i,j \rangle} {\bf S}_i \cdot {\bf S}_j,
\end{equation}
where ${\langle i,j \rangle}$ denote nearest-neighbor (NN) sites on a
hypercubic lattice, and ${\bf S}_i$ the spin operator [of amplitude
${{\bf S}_i}^2=S(S+1)$] on site $i$. While long-range
order in the ground state of Eq.~\eqref{eq:H} is prohibited in
one dimension (1D)~\cite{BetheMerminWagner}, it can be shown
rigorously~\cite{LRO} that the system displays long-range order in $d=2$
for all $S\geq 1$. For $S=1/2$, there is no exact proof but the
numerical computations~\cite{NumLRO,SandvikEvertz} leave no doubt that
this is also the case. In this paper, we deal only with this extreme
quantum case $S=1/2$ and assume an even number $N$ of spins.

The Hamiltonian Eq.~\eqref{eq:H} exhibits SU(2) symmetry, as it
commutes with the total spin operator
${\bf S}_T=\sum_{i=1}^N {\bf  S}_i$.
The total spin quantum number $S_T$ is defined by the
eigenvalue $S_T (S_T +1)$ of the operator ${{\bf S}_T}^2$. 
For AF interactions, the ground state
for finite $N$ of the Hamiltonian Eq.~\eqref{eq:H} on the hypercubic
lattice can be proven~\cite{LiebMattis} to be a total singlet $S_T=0$.
Natural objects to describe such a singlet state are 
{\it valence bonds} (VBs), which are intrinsically non-magnetic
and antisymmetric objects, with two spins $1/2$ coupled into a
singlet. This notion was introduced in the early days of quantum
mechanics~\cite{old} and has regained interest thanks to the proposal
that wave functions composed of VBs [``resonating valence bond (RVB)
wave functions''] could
describe a quantum spin-liquid phase without any magnetic order,
for Heisenberg models on frustrated lattices~\cite{Anderson}.
The idea is that local (typically NN)
superpositions of VBs can effectively accommodate the constraint
imposed by frustration in the building blocks of such lattices and
lower the energy of the full system. While the original
suggestion~\cite{Anderson} of this scenario taking place on the
triangular lattice is now abandoned (since this lattice supports AF
long-range order~\cite{TriangularLRO}), it stimulated many
investigations which led to significant progresses in
quantum magnetism.
In fact, the existence of the RVB phase is established in several
quantum dimer models~\cite{MoessnerSondhi,MisguichSerbanPasquier},
which are effective models of quantum
antiferromagnets. Furthermore, the usefulness of VB wave functions has 
been demonstrated for the understanding of the low-energy physics of Heisenberg AF models on other frustrated
lattices, such as kagome~\cite{kagome} or frustrated square
lattices~\cite{Mambrini06,LouSandvik,Zhang13}.
These developments have confirmed the importance of the
VB picture in understanding quantum antiferromagnets.

Much of the studies of quantum antiferromagnets based on the
VB picture is aimed at quantum spin liquids without
a long-range antiferromagnetic order.
However, it should be noted that the set of VB states,
once including longer-range VBs beyond nearest neighbors,
is {\it overcomplete} in the total singlet sector~\cite{old,old2}.
Thus, any singlet state can be represented as a superposition
of VB basis states. The finite-size ground state of
the antiferromagnetic Hamiltonian on the hypercubic lattice
in $d \geq 2$ dimensions is no exception, despite
its long-range antiferromagnetic order.
This somewhat counterintuitive fact is indeed consistent with
the existence of symmetry-broken ground states in the thermodynamic
limit, thanks to the asymptotic degeneracy in the thermodynamic
limit $N\to \infty$ of the ``Anderson tower of states'' of nonzero
total spin $S_T$ with the ground state~\cite{AndersonBook}.

An antiferromagnet is called a bipartite antiferromagnet
if the sites can be grouped into two sublattices, and
the Heisenberg exchange interactions exist only between
two sites belonging to different sublattices.
Examples include the antiferromagnets on a hypercubic
lattice with only nearest-neighbor interactions.
It is then natural to consider a restricted set of VB basis
states, bipartite VB basis states, in which only VBs
connecting different sublattices are allowed~\cite{BeachSandvik}.
Even the bipartite VB basis, if longer-range VBs are included,
is overcomplete in the total singlet sector (see {\it e.g.} Ref.~\onlinecite{BeachSandvik} or Ref.~\onlinecite{Mambrini})
and thus can represent any singlet ground state
of an $S=1/2$ antiferromagnet.
In fact, 
in a pioneering work by Liang, Dou\c{c}ot and Anderson
(LDA)~\cite{LDA}, the long-range-ordered AF ground state of the Heisenberg
antiferromagnet on the square lattice was described
in terms of bipartite VB basis states, including
bipartite VBs of arbitrary length. In practice, this
was done by defining a so-called amplitude product state, where the weight
of every valence bond state is factorized into weights $h({\bf r})$ coming
from contributing valence bonds of length ${\bf r}$. The original results of
LDA pointed to a minimal variational energy for the square lattice
Heisenberg model reached for a power law decay $h(r)\propto r^{-\alpha}$,
with $\alpha\approx4$.

The VB basis provides a new perspective for understanding quantum
antiferromagnets, which is complementary to the more traditional
$S^z$ basis. As an indication of its significance,
a quantum Monte Carlo (QMC) algorithm based on the VB basis
was proposed~\cite{Sandvik05} and proved to be useful.
The more recent numerical results from the VB-based
QMC simulation~\cite{Sandvik05},
obtained without assuming the power-law decay,
imply $\alpha=3$ rather than $\alpha \sim 4$ suggested by LDA.
This power-law decay and the value $\alpha=3$ can be understood
through a master
equation~\cite{BeachMaster} or other mean-field~\cite{Beach,Wegner}
approaches, which moreover predict $\alpha=d+1$ to be the best
variational ansatz for the ground state of Eq.~\eqref{eq:H}
in dimension $d$, within amplitude product states.
These previous works were explicitly assuming a
certain amplitude product variational ansatz for the antiferromagnetic
ground state~\cite{note-beach}. However, one could instead also ask
for the length distribution $P({\bf r})$ of valence bonds in the
{\it  real} ground state of Heisenberg antiferromagnets. This question
may at first glance seem ill defined since the expression of singlet
states in terms of valence bonds is in general
non-unique. Fortunately, the occupation number
of {\it bipartite valence bonds} is nevertheless still
well defined~\cite{Mambrini,Alet10}.

In this paper, we compute by analytical means the distribution of VB
lengths $P({\bf r})$ for antiferromagnetic Heisenberg models on
hypercubic lattices in dimension $d$.
Somewhat surprisingly, the question of VB distribution
can be reduced to a problem of
{\it boundary field theory}.~\cite{YellowPages,BlumenhagenPlauschinn}
Boundary field theory has been found useful, not only
in boundary critical phenomena and quantum impurity problems,
but also in recent topics such as
entanglement entropy.~\cite{FradkinMoore,
MO-EE-BCFT,JacobsenSaleur,StephanMisguichPasquier}
The VB distribution may be added to the list of applications
of boundary field theory.
The key in the relation is the identity~\cite{Alet10} between
the bipartite VB distribution and an overlap with
the reference state, which is naturally related to
boundary correlation functions.

In principle, any field theory, which is an appropriate
effective theory for the bulk, can be used to calculate
the bipartite VB distribution with a certain boundary condition.
The ground state and low-energy excitations
of the Heisenberg antiferromagnet on
the hypercubic lattice in $d \geq 2$ spatial dimensions
are believed to be described by the O(3) non-linear $\sigma$ model.
In fact, in the ground state, the SU(2) symmetry is spontaneously
broken in the non-linear $\sigma$ model in  $d \geq 2$, as is the
case in the Heisenberg antiferromagnet on the hypercubic lattice.
The effective field theory in the low-energy limit
is then reduced to the field theory of two massless
Nambu-Goldstone modes.
We will apply such a field theory with boundary, in order
to study the bipartite VB distribution in $d \geq 2$ dimensions.

On the other hand, in $d=1$, the physics is quite different as
there is no spontaneous breaking of the symmetry.
The system belongs to the universality class of
Tomonaga-Luttinger liquids, which can be described by
a free-boson field theory in $d=1$ spatial dimension.
Again, together with an appropriate boundary condition,
it is applied to the bipartite VB distribution in $d=1$
dimension. 
As a result, in $d \geq 1$ dimensions,
we find a power-law dependence $P({\bf  r}) \propto |{\bf r}|^{-(d+1)}$,
confirming previous results based on
mean-field approximations~\cite{Wegner,Beach,BeachMaster} as well as a
previous numerical QMC
estimate~\cite{Sandvik05}. The analytical predictions are compared in
detail to the results of QMC simulations of the Heisenberg model in
dimension $d=1$, 2 and 3.

While ``free-boson field theory'' is a common ingredient for the cases
$d=1$ and $d \geq 2$, the correspondence between the field
theory and the quantum antiferromagnet is quite different, reflecting
the very different physics.
In fact, the difference between $d=1$ and $d \geq 2$
becomes apparent in the {\it correlations} between VBs, as follows.
In the wave function studied by LDA, it is assumed
that the weight of each VB configuration can be factorized in products of
weights  carried by each VB individually: 
\begin{eqnarray}
\label{eq:LDA} | \psistate_{\rm LDA} \rangle = \sum_c
w_c | c \rangle,\\
{\rm with \quad } w_c = \prod_{(i,j) \in c} h_{ij}, \nonumber
\end{eqnarray}
where $c$ denotes a VB configuration composed of different VBs $(i,j)$
formed between spins at sites $i$ and $j$. The amplitude $h_{ij}$ is
often chosen as depending on the distance $r_{ij}$ between $i$ and
$j$, {\it i.e.}, the ``length'' of the VB $(i,j)$. 
The factorization of VB amplitudes means the absence of
correlation among VBs.
It is indeed a strong assumption in the ansatz
LDA wave function, and very few works~\cite{HuseElser} to our best
knowledge went beyond it until recently~\cite{BeyondLDA,Lin12,Zhang13}. In fact,
it is very natural to ask whether this assumption is correct or can be
justified for realistic (short-range) Heisenberg models.

The boundary field theory approach allows us to
elucidate analytically the correlation among VBs.
We find that, whereas there are nontrivial correlations
between VB occupation numbers even at large distances
for $d=1$, no correlations are asymptotically present in $d\geq 2$. 
The analytical prediction is then carefully
verified via a numerical investigation with
QMC simulations.
This difference between $d=1$ and $d\geq 2$
is another manifestation of different physics,
in terms of the VB picture. Our results for $d\geq 2$ also justify the use of factorized
wave functions of the LDA type.

The plan of the paper is the following. In Sec.~\ref{sec:def}, we
review the notion of valence bond occupation and correlation in a
singlet wave function, giving a pedagogical derivation of a simple
formula to effectively count the ``average'' number of VBs shared by
two sites. In Sec.~\ref{sec:qmc}, we present a method to measure these quantities
 numerically using QMC algorithms in the VB basis, and
discuss in particular a specific improved algorithm for doing so. We
then consider the case of $d\geq 2$
Heisenberg antiferromagnets using a non-linear $\sigma$ model
description in Sec.~\ref{sec:sigma}, computing the VB length
distribution as well as correlations. We furthermore compare the
analytical predictions for the Heisenberg model on the square and
simple cubic lattices with QMC results.  In
Sec.~\ref{sec:bos}, we study the more complicated case (from the
analytical point of view) of $d=1$ with bosonization, showing the
presence of correlations between VBs. The numerical computations in
$d=1$ are simpler and allow for an exhaustive comparison with
bosonization results. We further discuss the implication of our
results in Sec.~\ref{sec:discussion} and give a conclusion.

\section{Definition}\label{sec:def}
\begin{figure*}[t]
\includegraphics[width=0.9\textwidth]{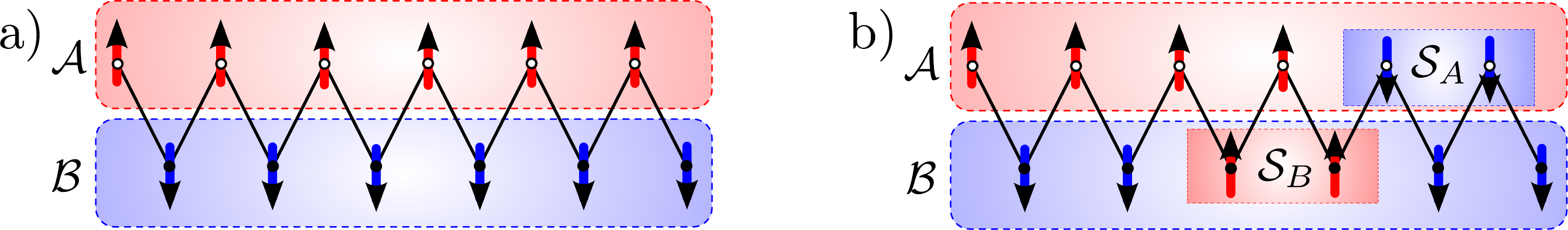}
\caption{(Color online) a) The reference N\'eel state (on a
  one-dimensional lattice) with all spins pointing up (down) on
  sublattice $A$ ($B$) has on overlap of $2^{-N/4}$ with all bipartite
  valence bond states, since each of the $N/2$ valence bonds
  contributes with a factor $1/\sqrt{2}$. b) Swapping the spins of the
  N\'eel state in region ${\cal S}_A$ with those in region ${\cal
    S}_B$ creates a reference state, which has a slightly different
  overlap with bipartite valence bond states. Every valence bond
  connecting ${\cal S}_A$ and ${\cal S}_B$ contributes a factor of
  $-1/\sqrt{2}$, whereas bipartite valence bonds, being only connected
  to one of the swapped regions, contribute a factor $0$. Depending on
  the size of the swapped regions, one can thus measure the
  single VB occupations ($|{\cal S}_A|=1=|{\cal S}_B|$),
  simultaneous occupations of two VBs
  ($|{\cal S}_A|=2=|{\cal S}_B|$), and others.}
\label{fig:occupation}
\end{figure*}

Consider a bipartite lattice, such that interactions in the
Hamiltonian Eq.~\eqref{eq:H} occur only between sublattice $A$ and
sublattice $B$. Every bipartite valence bond state
$\fullket{\varphi_\alpha}$ on such a lattice may be written as
\begin{equation}\label{eq:bipartiteVB}
\fullket{\varphi_\alpha} =
\fullket{(i_1,j_1)(i_2,j_2)\cdots(i_{N/2},j_{N/2})},
\end{equation}
with $i_k\in{\cal A}$ on sublattice $A$ and $j_k\in{\cal B}$ on
sublattice $B$. In this notation the pairs $(k,l)$ represent precisely
the bipartite valence bonds
\begin{equation}
\fullket{(k,l)} =
\frac{\fullket{\uparrow_k\downarrow_l}-\fullket{\downarrow_k\uparrow_l}}{\sqrt{2}},
\end{equation}
that form the state $\fullket{\varphi_\alpha}$. We may then define the
occupation number $n_{k,l}$ of a valence bond state as
\begin{equation}\label{eq:genoccupation}
n_{k,l}(\fullket{\varphi_\alpha}) =
\twopartdef{1}{(k,l)\in\fullket{\varphi_\alpha},}{0}{\mbox{otherwise}.}
\end{equation}
Note, that this definition is rather formal and not always very
useful. Using the N\'eel state
\begin{equation}\label{eq:neel}
\fullket{\mbox{N\'eel}} = \prod_{i\in{\cal
    A}}\fullket{\uparrow_i}\prod_{j\in{\cal B}}\fullket{\downarrow_j},
\end{equation}
an interesting alternative expression can be given in the form of a scalar
product~\cite{Alet10} :
\begin{equation}
\label{eq:nkldef}
n_{k,l}(\fullket{\varphi_\alpha}) =
-\frac{\left<\mbox{N\'eel}\middle|S_k^+S_l^-\middle|\varphi_\alpha\right>}{\left<\mbox{N\'eel}\middle|\varphi_\alpha\right>}.
\end{equation}
Therein, the operator $S_l^+S_k^-$ permutes the spins $k$ and $l$
within the N\'eel state $\fullket{\mbox{N\'eel}}$, as illustrated in
Fig.~\ref{fig:occupation}. Notice that here we swap only a single
spin ($|{\cal S}_A|=1=|{\cal S}_B|$), with ${\cal S}_A=\{k\}$ and
${\cal S}_B=\{l\}$. Obviously, a valence bond $(k,l)$ in
$\fullket{\varphi_\alpha}$ contributes a factor of $1/\sqrt{2}$ when
overlapped with $\fullbra{\uparrow_k\downarrow_l}$ and $-1/\sqrt{2}$
for the overlap with $\fullbra{\downarrow_k\uparrow_l}$. Hence, a
valence bond connecting ${\cal S}_A$ and ${\cal S}_B$ must yield a
factor of $1$ in Eq.~\eqref{eq:nkldef}, as required. On the other
hand, if there are valence bonds $(k,m)$ that connect ${\cal S}_A$ and
${\cal B}-{\cal S}_B$ (see Fig. \ref{fig:occupation}), then the
overlap with a triplet state vanishes
$\left<\downarrow_k\downarrow_m\middle|(k,m)\right>=0$, thus showing
the equivalence of Eq.~\eqref{eq:nkldef} with
Eq.~\eqref{eq:genoccupation}.

It is also straightforward to generalize Eq.~\eqref{eq:nkldef} to
arbitrary singlet states~\cite{Alet10,Mambrini}, which can always be
expressed as (albeit non-unique) superpositions of bipartite VB
states. Writing a given singlet state $|\psistate\rangle$ as
\begin{equation}
\fullket{\psistate} = \sum_i a_i \fullket{\varphi_i},
\label{eq.psi_in_vb}
\end{equation}
we can immediately generalize Eq.~\eqref{eq:nkldef} to
the ``average'' VB occupation number for $|\psistate\rangle$ as
\begin{equation}
\label{eq.nkl_psi}
\bar{n}_{(k,l)}(\fullket{\psistate}) =
-
\frac{\left< \protect\mbox{N\'eel} \middle| S_k^+S_l^- \middle|
\psistate \right>}{
\left< \protect\mbox{N\'eel} \middle| \psistate \right>}.
\end{equation}
Even though the expression Eq.~\eqref{eq.psi_in_vb}
of the singlet state in terms of VB basis states is not unique,
the VB occupation number Eq.~\eqref{eq.nkl_psi} is unique and
well defined.
Since every VB basis state has
the same overlap with the N\'eel state, that is, 
$\left< \protect\mbox{N\'eel} \middle| \varphi_i \right>
= 2^{-N/4}$, 
Eq.~\eqref{eq.nkl_psi} is thus an ``average'' defined with
respect to the coefficients $a_i$ themselves as the weight,
and not to their squares $|a_i|^2$.
If some of the coefficients $a_i$ can be negative,
the average Eq.~\eqref{eq.nkl_psi} may be ill defined.
However, for the ground state of a bipartite quantum antiferromagnet,
all the coefficients $a_i$ of the bipartite VB basis states
are non-negative upon a proper choice of the overall phase factor,
thanks to Marshall's sign rule~\cite{Marshall, LDA}.
Thus the average occupation number is always well-defined for
such a ground state and satisfies $0 \leq \bar{n}_{(k,l)} \leq 1$.
We emphasize that this only applies to
bipartite valence bonds~\cite{Mambrini,Alet10}.

We define the single VB distribution function
\begin{equation}
P(\mathbf{r}) \equiv \bar{n}_{(k,l)},
\end{equation}
where two sites
$k,l$ are separated by the vector $\mathbf{r}$.
As we consider translationally invariant systems,
$\bar{n}_{k,l}$
should not depend on the choice of the site $k$
and is a function of the separation ${\bf r}$ only.
Moreover, since the site $k$ must belong to one VB in any of the
VB basis states, we find the sum rule
\begin{equation}
\sum_{\mathbf{r}} P(\mathbf{r}) = 1. 
\label{eq.Pr_sumrule}
\end{equation}

This construction carries over immediately to the more general problem
of simultaneous occupation numbers of multiple VBs.
This can be done by choosing the set of sites ${\cal S}_A$ and
${\cal S}_B$ from each sublattice $A$ and $B$, so that
$|{\cal S}_A|=|{\cal S}_B|>1$.
Then, the ``number of VBs connecting ${\cal S}_A$ and ${\cal S}_B$''
can be obtained as
\begin{align}
\label{eq:gencorrelations}
\bar{n}_{{\cal S}_A,{\cal S}_B}(\fullket{\psistate}) & =
-\frac{\left<\mbox{N\'eel}\middle|\prod_{k\in{\cal S}_A}S_k^+\prod_{l\in{\cal S}_B}S_l^-\middle|\psistate\right>}{\left<\mbox{N\'eel}\middle|\psistate\right>}
\notag \\
 & = -\frac{\sum_i a_i \left<\mbox{N\'eel}\middle|\prod_{k\in{\cal S}_A}S_k^+\prod_{l\in{\cal S}_B}S_l^-\middle|\varphi_i\right>}{\sum_i a_i \left<\mbox{N\'eel}\middle|\varphi_i\right>}.
\end{align}
We note that, in this quantity,
one no longer sees which sites precisely in
${\cal S}_A$ and ${\cal S}_B$ are connected to each other.
For example, with the choice of ${\cal S}_A = \{k,l\}$
and ${\cal S}_B = \{m,n\}$, 
we measure the simultaneous occupation of two VBs connecting
the two sets.
This may be interpreted as a sum of the occupation numbers for 
the two possible VB configurations $\fullket{(k,m)(l,n)}$ and
$\fullket{(k,n)(l,m)}$:
\begin{align}
\bar{n}_{\{k,l\},\{m,n\}}(\fullket{\psistate}) & = 
\bar{n}_{(k,m)(l,n)}(\fullket{\psistate}) + 
\bar{n}_{(k,n)(l,m)}(\fullket{\psistate})
\notag \\
& =
-\frac{\left<\mbox{N\'eel}\middle|S_k^+ S_l^+ S_m^- S_n^- \middle|
\psistate\right>}{\left<\mbox{N\'eel}\middle|\psistate\right>}.
\label{eq.2VBocc}
\end{align}
We assert that, however, the occupation numbers
for the individual VB configurations
$\fullket{(k,m)(l,n)}$ or $\fullket{(k,n)(l,m)}$
are not well defined, and only the combination Eq.~\eqref{eq.2VBocc}
is uniquely defined.

\section{Measuring valence bond occupations and correlations in Quantum Monte Carlo simulations}\label{sec:qmc}

Measuring valence bond occupation numbers or correlations is
straightforward using Eq.~\eqref{eq:gencorrelations} if one has direct
access to the wave function studied, such as within exact
diagonalization or density matrix renormalization group (DMRG) calculations (see Ref.~\onlinecite{Alet10} for a more precise
discussion and an example of such computations). Since, however, we
would like to study asymptotic, long-distance, properties of $d>1$
quantum antiferromagnets on large systems, these methods are
not practical and we have to resort to stochastic QMC schemes.

\subsection{General statements on valence bond QMC}

Recent QMC state-of-the-art algorithms allow access to quantities in the valence bond language directly, by projecting out the ground state from some trial wave function. The trial wave function is usually chosen to be some special singlet state, such as the amplitude product state Eq.~\eqref{eq:LDA}, in order to achieve a significant reduction of computational cost~\cite{SandvikBeach}.

Initially, two such schemes have been devised~\cite{Sandvik05}, which are often referred to as single projection and double projection, respectively. The choice between these two schemes depends on the observable we want to measure. Observables that commute with the Hamiltonian can be measured within the single projection scheme, where some power $m$ of the Hamiltonian ${\cal H}$ is applied to a singlet trial state $\fullket{\psistate_{\rm in}}$,
\begin{equation}\label{eq:singleprojection}
\left<{\cal O}\right> = \frac{\fullbra{\mbox{R}}{\cal O}(C-{\cal H})^m\fullket{\psistate_{\rm in}}}{\fullbra{\mbox{R}}(C-{\cal H})^{m}\fullket{\psistate_{\rm in}}}.
\end{equation}
Herein, $\fullket{\mbox{R}}$ is a reference state, which is in principle arbitrary.
However, it turns out that the N\'eel state $\fullket{\mbox{\Neel}}$
is a particularly good choice, as
it has the same overlap with all singlet states. Furthermore, $C$ is
chosen to be the energy of the maximally polarized state
$\fullket{\uparrow\cdots\uparrow}$, such that the projection reveals
the ground state.

Notice, that such a scheme requires the ground state $(C-{\cal H})^{m}\fullket{\psistate_{\rm in}}$ to be an eigenstate of the observable ${\cal O}$. In cases where this does not apply we have to carry out the double projection scheme, where the ground state projection occurs on both sides,
\begin{equation}\label{eq:qmcobservable}
\left<{\cal O}\right> = \frac{\fullbra{\psistate_{\rm in}}(C-{\cal H})^m{\cal O}(C-{\cal H})^m\fullket{\psistate_{\rm in}}}{\fullbra{\psistate_{\rm in}}(C-{\cal H})^{2m}\fullket{\psistate_{\rm in}}}.
\end{equation}

Valence bond occupation, Eq.~\eqref{eq:gencorrelations}, is a rather peculiar
quantity, which measures properties of a given singlet state without being
an observable. There are however different ways to generate related observables, which can be measured within double projection, {\it e.g.},
\begin{align}
{\cal O}_{k,l} & = S_k^-S_l^+\left|\mbox{N\'eel}\middle>\middle<\mbox{N\'eel}\right|S_k^+S_l^-,\\
{\cal P}_{k,l} & = -\frac{1}{2}\left( \left|\mbox{N\'eel}\middle>\middle<\mbox{N\'eel}\right|S_k^+S_l^- + S_k^-S_l^+\left|\mbox{N\'eel}\middle>\middle<\mbox{N\'eel}\right| \right).
\end{align}
Unfortunately, it turns out that simulations converge rather badly when carrying out a double projection scheme with these observables. This is due to the fact that the appropriate estimator fluctuates exponentially, like the one for fidelity~\cite{Schwandt09} and R\'enyi entropies~\cite{Hastings10} and in contrast to the energy estimator and other quantum mechanical observables~\cite{Sandvik05}.

On the other hand, one notices that Eq.~\eqref{eq:gencorrelations} looks
very similar to the single projection expression
Eq.~\eqref{eq:singleprojection}, suggesting an implementation of this last
scheme. Indeed valence bond occupations (through the related valence bond
entanglement entropy~\cite{vbee}) have already been measured with the
single projection method using a local update algorithm. However, whereas for the double projection method very efficient loop updates are known~\cite{SandvikEvertz}, this seems not to be the case for the single projection algorithm. We devote the next section to the description of such a method.

\subsection{Single projection QMC algorithm with loop updates}

In order to carry out a single projection scheme, we first rewrite expression Eq.~\eqref{eq:singleprojection} and decompose the projection step as a sum over all possible contributing bond operator strings~\cite{Sandvik05,SandvikEvertz}
\begin{equation}
(C-{\cal H})^m=\sum_r {P}_r^{(m)}.
\end{equation}
Every one of such operator strings is a product of bond singlet projectors,
\begin{equation}
{P}_r^{(m)} = {\cal P}_{b_1}^{(s=0)}\cdots {\cal P}_{b_m}^{(s=0)},
\end{equation}
and $r=(b_1,\ldots,b_m)$ is a multi-index that is summed over. Notice, that
the singlet projector ${\cal P}_{b}^{(s=0)}$ on a given bond $b=\, \langle i,j\rangle$ between sites $i$ and $j$ is related to the spin-spin interaction by
\begin{equation}
{\cal P}_{\langle i,j\rangle}^{(s=0)} = \frac{1}{4} - {\bf S}_i \cdot {\bf S}_j.
\end{equation}

\begin{figure}[t!]
\includegraphics[width=0.95\columnwidth]{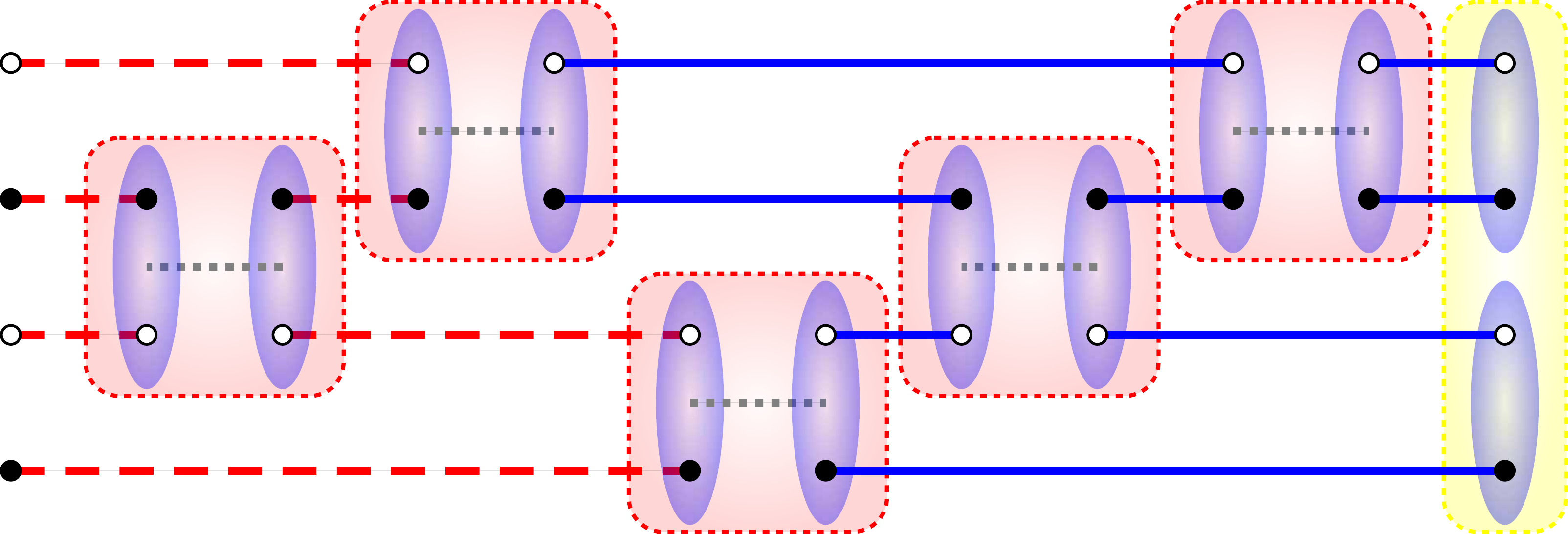}
\caption{(Color online) Schematic representation of the expression
  $\fullbra{\uparrow\downarrow\uparrow\downarrow}{\cal P}_{\langle
    2,3\rangle}^{(s=0)}{\cal P}_{\langle 1,2\rangle}^{(s=0)}{\cal
    P}_{\langle 3,4\rangle}^{(s=0)}{\cal P}_{\langle 2,3
    \rangle}^{(s=0)}{\cal P}_{\langle 1,2\rangle
  }^{(s=0)}\fullket{(1,2)(3,4)}$, residing on $N=4$ sites (drawn
  vertically). Herein, the singlet projectors (red shaded squares) are
  intentionally represented by two valence bonds. Notice that the N\'eel
  state on the left (represented by black and white circles) does not
  close the red (dashed) loops, whereas the blue (continuous) loops
  are closed intrinsically or by the valence bond state (yellow shaded
  rectangle) on the right. It is very straightforward to see, that the
  above expression can be evaluated by Sutherland's~\cite{Sutherland}
  overlap rule $2^{N_\bigcirc-N_v/2}$, where $N_\bigcirc$ denotes the
  number of \textit{closed} loops and $N_v$ the \textit{total} number
  of valence bonds. Here we count $N_\bigcirc=2$ closed loops and
  $N_v=5\times2+2=12$ valence bonds.}
\label{fig:loopgraph}
\end{figure}

With these notations and noting that the singlet trial state can be decomposed as a superposition of bipartite VB states $\fullket{\psistate_{\rm in}} = \sum_i a_i \fullket{\varphi_i}$,
Eq.~\eqref{eq:singleprojection} reads
\begin{equation}\label{eq:newsingleprojection}
\left<{\cal O}\right> = \frac{\sum_{r,i} a_i
  \fullbra{\mbox{\Neel}}{\cal O} {P}_r^{(m)}
  \fullket{\varphi_i}}{\sum_{r,i} a_i
  \fullbra{\mbox{\Neel}}{P}_r^{(m)}\fullket{\varphi_i}},
\end{equation}
where $\fullket{\mbox{\Neel}}$ is used as the reference state and the sums over $r$ and $i$ can be sampled stochastically. The way
this is done is illustrated in Fig.~\ref{fig:loopgraph}, where an
example bond operator string is explicitly written out. We use the
following updates in order to sample the configuration space of
different bond operator positions and valence bond states.

\textit{Loop updates.} The loop updates~\cite{SandvikEvertz} can be applied
here with a slight modification, where loops passing through the N\'eel
state cannot be flipped as in the double projection
scheme~\cite{SandvikEvertz}. This can be understood in two different ways:
First, in the original scheme loop flips correspond to a flip of
underlying spin configurations. Whereas every valence bond and every singlet
projector contains two such configurations, the N\'eel state is a single spin configuration. Therefore, loops through the N\'eel state exist only in one flavor and cannot be flipped, in contrast to all other loops that exist in two flavors.

Another way of seeing this is by looking at the interpretation of
Sutherland's overlap rule~\cite{Sutherland} described in Fig.~\ref{fig:loopgraph}, which
deals only with closed loops but not with open ones. The loop algorithm aims precisely to mimic the varying contributions from the number of loops, by rewriting the term $2^{N_\bigcirc}$ as a sum over different configurations. Allowing each closed loop to take two different flavors generates exactly the required sum, whose stochastic sampling is computationally cheaper than counting the number of closed loops in a given overlap graph.

\begin{figure}[ht]
\includegraphics[width=0.99\columnwidth]{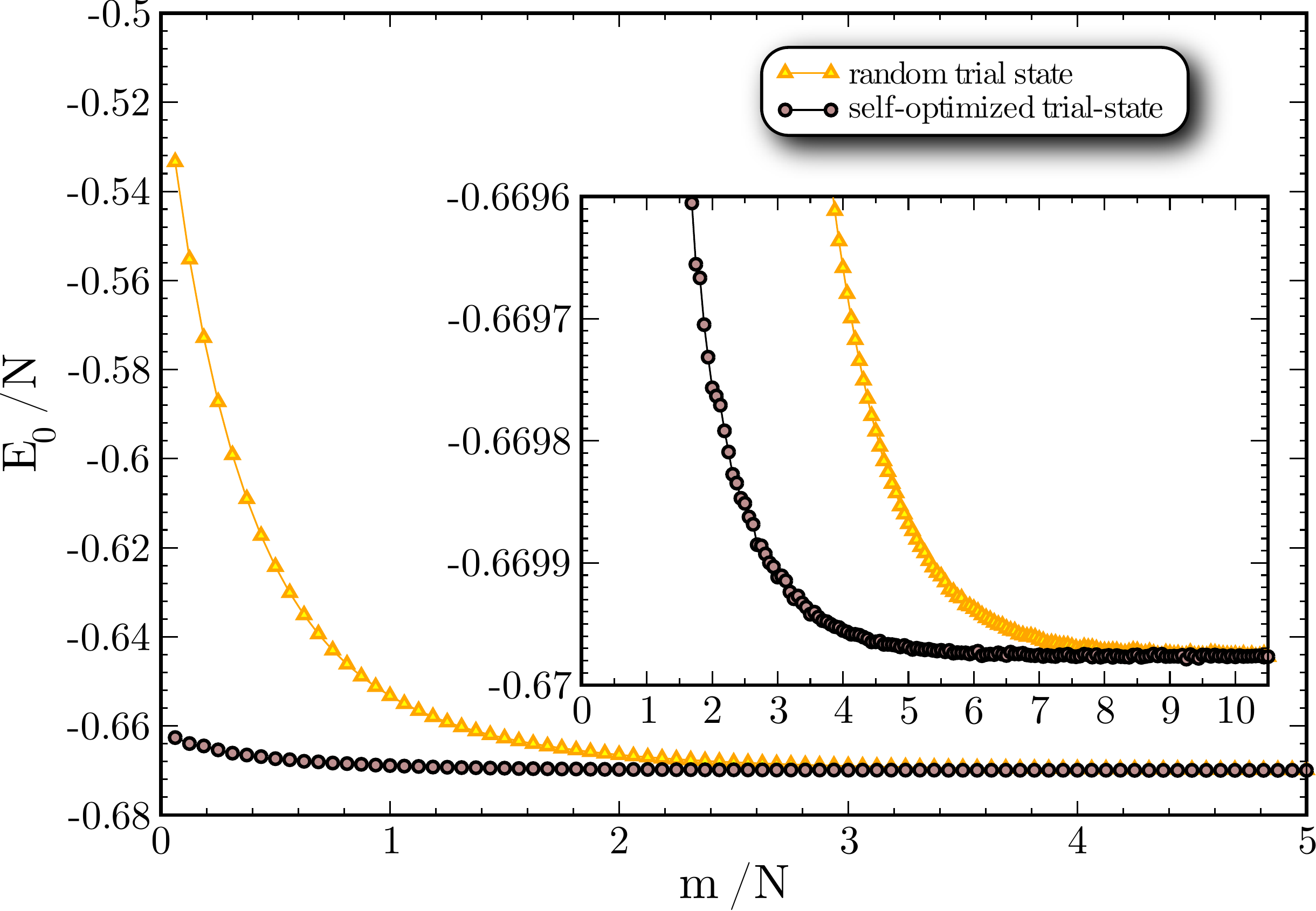}
\caption{(Color online) Convergence of the ground state energy of a 2D square lattice system with $N=16^2$ sites versus the projection length $m$ in a single projection scheme. Inset is a zoom on the low-energy part. For the self-optimized amplitude product state a significant convergence speed up is achieved (error bars are smaller than the point sizes).}
\label{fig:selftrial}
\end{figure}

As a result, we keep the flavor of the red (dashed) loops in
Fig.~\ref{fig:loopgraph} constant, while we flip every blue (continuous)
loop with a probability of $50\%$ between the two flavors. As in the
original scheme~\cite{SandvikEvertz}, the different flavors represent a
constraint for all other updates, which can be carried out in completely the
same manner as in the original work. This includes \textit{operator
  updates}, which are used to move the bond operators into different
positions of the overlap graph, as well as \textit{state updates}, in order
to sample the different valence bond states
$\fullket{\varphi_i}$. Self-optimized amplitude product states~\cite{SandvikBeach} turn out to be an advantageous choice for the single projection scheme as well, as can be seen in Fig.~\ref{fig:selftrial}. We observe the same benefit for quantities such as Eq.~\eqref{eq:gencorrelations}. The probabilities of both such updates remain the same as for the double projection scheme~\cite{SandvikEvertz}.
  
\textit{Measurements.} As the sampling weight $W_{r,i}^{(m)}$ is simply given by
\begin{equation}
W_{r,i}^{(m)} = \frac{a_i \fullbra{\mbox{\Neel}} {P}_r^{(m)} \fullket{\varphi_i}}{\sum_{r,i} a_i \fullbra{\mbox{\Neel}}{P}_r^{(m)}\fullket{\varphi_i}} \sim a_i,
\end{equation}
where the remaining factors keep constant during the simulation, the estimator is given by
\begin{equation}
O_{r,i}^{(m)} = \frac{\fullbra{\mbox{\Neel}}{\cal O} {P}_r^{(m)} \fullket{\varphi_i}}{\fullbra{\mbox{\Neel}}{P}_r^{(m)}\fullket{\varphi_i}}.
\end{equation}
Notice that the evaluation of this last expression is very simple,
since the denominator is a constant and cancels most of the terms in
the numerator. At the end one only has to determine the projected
valence bond state ${P}_r^{(m)} \fullket{\varphi_i}$, which can
directly be read out of the overlap graph (Fig. \ref{fig:loopgraph}) in the following way: Two sites form a valence bond in ${P}_r^{(m)} \fullket{\varphi_i}$ if they are linked together by an open loop in Fig. \ref{fig:loopgraph}. Graphically, one just needs to start from any initial site on the N\'eel state side and follow the open loop until it reaches a second site on the N\'eel state : The initial and second sites are then coupled in a valence bond in ${P}_r^{(m)} \fullket{\varphi_i}$. The operator ${\cal O}$ has then to be
inserted between this resulting valence bond state and the N\'eel
state itself.

\subsection{Application to valence bond occupations and correlations}

In the case of valence bond occupations and valence bond correlations the
operator ${\cal O}$ is replaced by $-S_k^+S_l^-$, which amounts to measure
$1$, if the projected valence bond state ${P}_r^{(m)} \fullket{\varphi_i}$
contains a valence bond $(k,l)$, and $0$ if not. In practice, we measure a
histogram of all valence bonds in ${P}_r^{(m)} \fullket{\varphi_i}$, where
we record the length $(x,y)$ of the valence bonds in case of valence bond
occupations, or their lengths together with their separation distance in case of valence bond correlations.

Whereas in one dimension it is possible to collect all data for every
single valence bond, in higher dimensions we will restrict ourselves to
measuring only specific configurations, as the amount of data would
otherwise not be manageable.

\section{Asymptotic behavior of VB occupation numbers in $d \geq 2$}
\label{sec:2d}

\subsection{Boundary field theory approach}
\label{sec:sigma}

Let us consider the statistics of VB occupations Eq.~\eqref{eq:gencorrelations},
in a field theory framework.
We note that, when $|\psistate\rangle$ is a ground state,
the average VB occupation number between sites
$k$ and $l$ [Eq.~\eqref{eq.nkl_psi}] for example is given as
\begin{equation}
\bar{n}_{(k,l)}(\fullket{\psistate}) = \lim_{\tau \rightarrow \infty}
- \frac{\langle \mbox{\Neel} | S^-_k S^+_l e^{-\tau \calH} 
| \psistate_{\rm in} \rangle }{
\langle \mbox{\Neel} | e^{-\tau \calH} | \psistate_{\rm in} \rangle },
\label{eq.vb_as_bcf}
\end{equation}
where $| \psistate_{\rm in} \rangle$ is the initial state, which
is arbitrary as long as the overlap
$\langle \psistate | \psistate_{\rm in} \rangle$ does not vanish.
This can be interpreted as a correlation function
$\langle S^-_i S^+_j \rangle$, in the presence of a boundary
described by the ``boundary state''
$| \mbox{\Neel} \rangle$~\cite{YellowPages,BlumenhagenPlauschinn}.
The boundary condition in the field theory,
which exactly matches the {\Neel } state 
$| \mbox{\Neel} \rangle$ as the boundary state, is unknown.

Nevertheless, the large-distance asymptotic behavior of
boundary correlation functions is governed by fixed points
of boundary renormalization group flow. Those boundary fixed points correspond to
conformally invariant boundary conditions.
Thus, as long as we are concerned with the
large-distance asymptotic behavior of the VB distribution,
we can replace the boundary condition with the
conformally invariant one corresponding
to the infrared fixed point for the boundary RG flow starting
from $|\mbox{\Neel} \rangle$.
In general, tracking the boundary RG flow is a difficult problem.
However, in the circumstances we will consider,
we can find a natural candidate for such a boundary condition.
This approach is generally applicable to the
evaluation of the amplitude (inner product) between a reference
state and the groundstate of a quantum many-body system in
any dimension, at least in principle.
We note that this approach indeed has been
employed~\cite{JacobsenSaleur,StephanMisguichPasquier}
in $d=1$ dimension, for problems closely related to
(but somewhat different from) the present one.
Here we demonstrate that it is also useful in $d \geq 2$,
and we will come back to $d=1$ in Sec.~\ref{sec:bos}.

The low-energy/large-distance effective theory of the
Heisenberg antiferromagnet on the hypercubic lattice
in $d\geq 2$ dimensions is the O(3) nonlinear $\sigma$ model.
It is defined by the Lagrangian density
\begin{equation}
 \calL = \frac{1}{2g} [\partial_\mu \vec{m}(\mathbf{r},\tau)]^2,
\label{NLsigma}
\end{equation}
where $\vec{m}$ is a three-dimensional vector
subject to the constraint
\begin{equation}
 \vec{m}^2 = 1.
\end{equation}
The field $\vec{m}$ describes the staggered component of the
original spin operator:
\begin{equation}
 \vec{S}_{\mathbf{r}} \sim
(-1)^{s(\mathbf{r})} S \vec{m}(\mathbf{r}) + \cdots .
\label{S_in_n}
\end{equation}

As we have discussed, the ground state of the
Heisenberg antiferromagnet in a $d \geq 2$ hypercubic
lattice generally displays a long-range antiferromagnetic
order. This also implies the spontaneous breaking
of the O(3) symmetry in the corresponding
non-linear $\sigma$ model, and
the field $\vec{m}$ is aligned to a particular direction.
Without losing generality, let us assume that
$\vec{m}$ is aligned in the $z$ direction.
The small fluctuation around the ``vacuum'' can be
described by the two components $m^x$ and $m^y$, so that
\begin{equation}
 m^z = \sqrt{1 - (m^x)^2 - (m^y)^2} .
\label{nz_in_nxy}
\end{equation}
The Lagrangian density Eq.~\eqref{NLsigma} is now written as
\begin{equation}
 \calL \sim \frac{1}{2g}
\left[ (\partial_\mu m^x)^2 + (\partial_\mu m^y)^2 \right],
\label{NGboson}
\end{equation}
where higher-order terms in $m^x$ and $m^y$
which come from the expansion of Eq.~\eqref{nz_in_nxy}
are neglected.
Physically, Eq.~\eqref{NGboson} represents the
theory of two Nambu-Goldstone modes which arise
due to the spontaneous symmetry breaking.
The ignored higher-order terms correspond to
interaction between Nambu-Goldstone bosons.
In the present case ($d \geq 2$ and $T=0$),
the interactions are irrelevant and
the theory of free Nambu-Goldstone bosons~\eqref{NGboson}
is asymptotically exact in the large-distance limit,
in the symmetry-broken phase.

While, in general,
the direction of ordering in the reference {\Neel } state
can be arbitrary in Eq.~\eqref{eq.vb_as_bcf},
in the present case we have to choose the reference {\Neel }
state with the same ordering direction as
in the ground state $|\psistate\rangle$.
Otherwise, the overlap would vanish.
Thus, here we take the reference {\Neel } state with
the ordering in the $z$ direction and use
Eq.~\eqref{eq.vb_as_bcf}.

In terms of the free Nambu-Goldstone boson field
theory Eq.~\eqref{NGboson}, Eq.~\eqref{eq.vb_as_bcf} is proportional to
\begin{equation}
 \langle \mbox{\Neel}_z | m^x(0) m^x(\mathbf{r}) | \psistate \rangle
\label{vboverlap-nxnx}
\end{equation}
(a similar term with $m^y$ gives the same contribution)~\cite{note1}. In the {\Neel } state, the spins are completely aligned
antiferromagnetically along the $z$ axis.
Considering Eq.~\eqref{S_in_n}, it is natural to
expect that (the infrared limit of) the boundary
condition corresponding to $| \mbox{\Neel}_z \rangle$
is the Dirichlet boundary condition for $m^x$ and $m^y$:
\begin{equation}
 m^x(\mathbf{r},0)  = m^y(\mathbf{r},0) = 0,
\label{nxnyD}
\end{equation}
imposing $\vec{m} = (0,0,1)$, which corresponds to
the perfect \Neel{} order in the $z$ direction,
at the boundary $\tau = 0$.

Thus the valence bond occupation number, in the
large-distance limit, is proportional to
the correlation function $\langle m^x m^x \rangle$
along the boundary with the Dirichlet boundary condition.
This correlation function appears to vanish, because
$m^x$ is set to zero on the boundary.
However, in adopting the boundary picture in the continuum limit, the
operator $m^x$ does not have to be exactly at the boundary; generally
we expect that they would be rather located within a short distance
$\epsilon$, of the order of the ultraviolet cutoff, from the boundary.
Namely, we postulate that
\begin{equation}
 \bar{n}_{\mathbf{0} \mathbf{r}} \propto
 \langle m^x(\mathbf{0}, - \epsilon) m^x(\mathbf{r}, - \epsilon) \rangle_D,
\label{nxnx-split}
\end{equation}
where $\langle \cdot \rangle_D$ is the expectation value
in the presence of the boundary at $\tau=0$, where
the Dirichlet boundary condition Eq.~\eqref{nxnyD} is imposed.
A similar trick, known as ``point-splitting technique''
is often used in applications of field theory.

Within the free Nambu-Goldstone boson theory Eq.~\eqref{NGboson},
correlation functions with the Dirichlet boundary condition
can be conveniently calculated by the method of mirror images.
For example,
Eq.~\eqref{nxnx-split} can be written as
\begin{align}
\bar{n}_{\mathbf{0} \mathbf{r}} & \propto
\langle m^x(\mathbf{0}, - \epsilon) m^x(\mathbf{r}, - \epsilon) \rangle_D
\notag \\
& =
\langle
\left[m^x(\mathbf{0}, \epsilon) - m^x(\mathbf{0}, - \epsilon)\right]
\left[m^x(\mathbf{r}, \epsilon) - m^x(\mathbf{r}, - \epsilon)\right]
\rangle,
\end{align}
in terms of the correlation functions denoted by
$\langle \cdots \rangle$ in the infinite plane.
For a small $\epsilon$, it can be expanded as
\begin{align}
\bar{n}_{\mathbf{0} \mathbf{r}} & \propto
4 \epsilon^2
\frac{\partial}{\partial \tau}
\frac{\partial}{\partial \tau'}
\langle m^x(\mathbf{0},\tau) m^x(\mathbf{r},\tau') \rangle
\big|_{\tau = \tau' = 0}
\notag \\
&= 
\const
\frac{\partial}{\partial \tau}
\frac{\partial}{\partial \tau'}
\frac{1}{\left[ \mathbf{r}^2 + (\tau - \tau')^2 \right]^{\frac{d-1}{2}}}
\big|_{\tau = \tau' = 0}
\notag \\
&=
\const \frac{1}{|\mathbf{r}|^{d+1}} .
\label{VBdist-d-dim}
\end{align}
Thus we have derived the power law length distribution
of VBs, in agreement with several earlier works~\cite{Wegner,Beach,BeachMaster}.
The present approach clarifies universality and
asymptotic exactness of this power law in quantum Heisenberg
antiferromagnets. 
Moreover, it enables us to discuss correlations among VBs
by analyzing the simultaneous occupation number of multiple VBs.
As the simplest among such cases, let us discuss
the simultaneous occupation number of two VBs
as introduced in Eq.~\eqref{eq.2VBocc},
and whether there is any correlation between two VBs.
Eq.~\eqref{eq.2VBocc} can be written in terms of
a four-point correlation function of
$m^x$ and $m^y$, through Eq.~\eqref{S_in_n}.
Within the free Nambu-Goldsone boson theory Eq.~\eqref{NGboson},
thanks to Wick's theorem,
the four-point correlation function is given in terms
of products of two-point correlation functions.
We thus find
\begin{align}
\bar{n}_{\{k_1,k_2\},\{l_1,l_2\}} & =
\langle m^x_{k_1} m^x_{l_1} \rangle
\langle m^x_{k_2} m^x_{l_2} \rangle
\notag \\
& \phantom{\langle m^x_{k_1} m^x_{l_1} \rangle
\langle m^x_{k_2} m^x_{l_2} \rangle}
+\langle m^x_{k_1} m^x_{l_2} \rangle
\langle m^x_{k_2} m^x_{l_1} \rangle
\notag \\
& =
\bar{n}_{(k_1,l_1)} \bar{n}_{(k_2,l_2)} +
\bar{n}_{(k_1,l_2)} \bar{n}_{(k_2,l_1)}
\notag \\
& \propto
\frac{1}{{r_{11}}^{d+1}} \frac{1}{{r_{22}}^{d+1}} 
+
\frac{1}{{r_{12}}^{d+1}} \frac{1}{{r_{21}}^{d+1}} ,
\end{align}
where $r_{mn}$ is the distance between sites $k_m$ and $l_n$.
Namely, it is given simply by a sum of products of the
average occupation numbers of the individual VBs.
This implies that there is in fact {\em no correlation}
in valence bond occupations between different pairs of sites,
and occupation of each pair occurs
{\em independently}.
This conclusion extends straightforwardly to correlations
among occupations of an arbitrary number of bonds.
It should be noted that the present field theory
describes the asymptotic behavior only in the large-distance limit.
The above result does not exclude correlations
in a quantum antiferromagnet at short length scales. 

\subsection{$d=2$ QMC results}\label{sec:QMC2d}
\begin{figure}[ht]
\includegraphics[width=0.99\columnwidth]{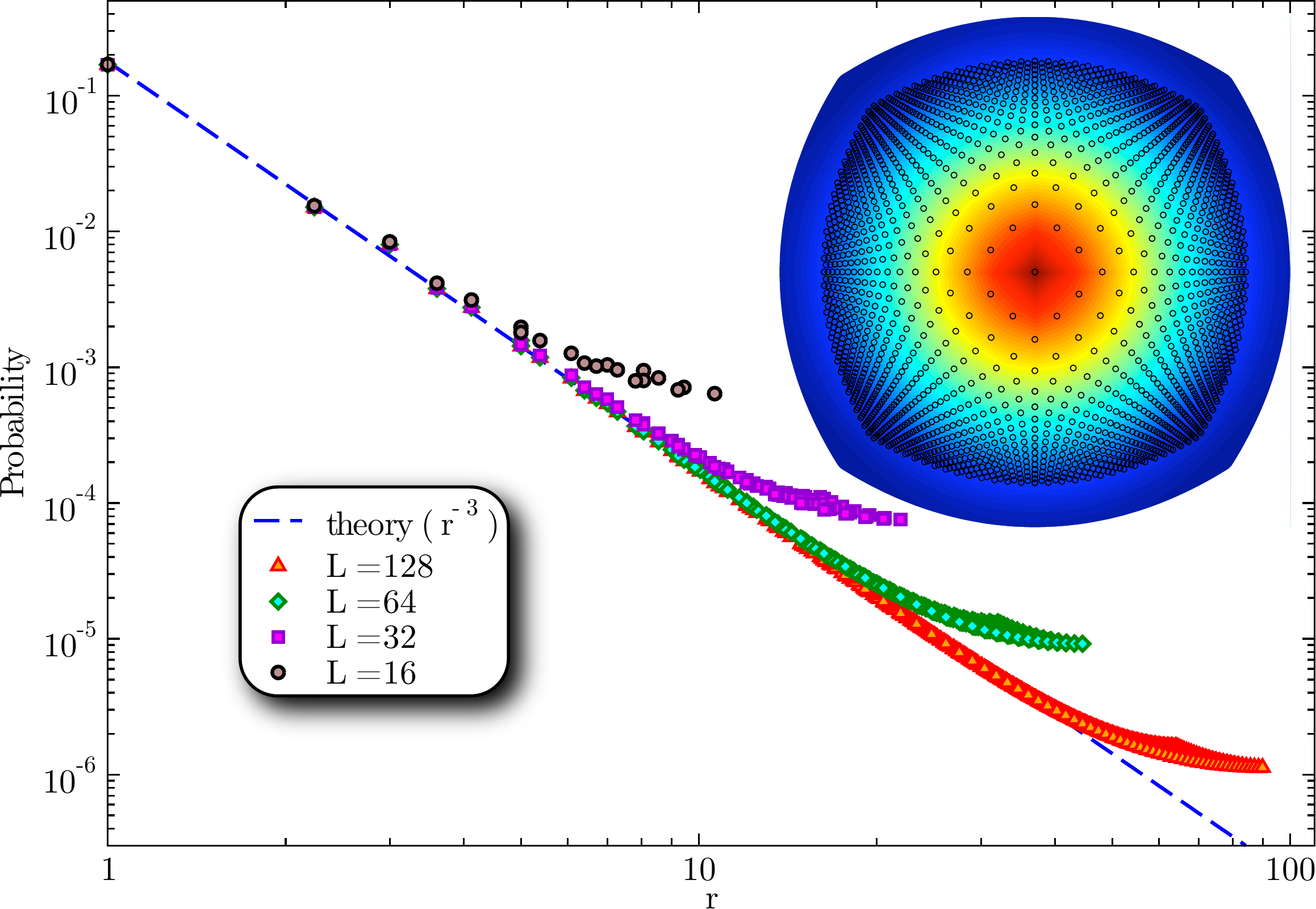}
\caption{(Color online) Length distribution $P(r)$ of valence bonds of
  length $r=\sqrt{x^2+y^2}$ (in all orientations) on different
  clusters with $N=L^2$ sites. The dashed blue line is a guide to the
  eyes, showing the expected $r^{-3}$ behavior from field theory. This
  scaling is observed as long as valence bonds are significantly shorter than
  the finite-size lattice sample. For very long valence bonds, the data exhibit a
  small upturn above this power law, as well as a slight anisotropy,
  due to the periodic boundary conditions. Data have been produced for
  a projection length of $m/N=16$ for all clusters except the biggest
  one, for which a projection length of $m/N=32$ was chosen
  (convergence was checked). Error bars are smaller than
  the point sizes.  Inset: Valence bond distribution $\log
  P(\log(r)\cos\theta,\log(r)\sin\theta)$ in logarithmic polar
  coordinates with logarithmic color scheme for $N=128^2$ in the
  region $[-L/2,L/2]^2$. The data grid is illustrated by small black
  circles in the range $[-L/4,L/4]^2$, whereas the colored mesh
  represents interpolated data. Outside this range the data grid is
  very dense and is therefore omitted. For short valence bonds the
  distribution appears to be rather isotropic, in agreement with the
  principal figure, whereas the boundary effects of the finite size
  sample can be observed for very long valence bonds.}
\label{fig:prVBD2d}
\end{figure}

Here we present the numerical results on the VB distribution
for the $S=1/2$ Heisenberg antiferromagnet on a $d=2$ square lattice.
One first issue to check is whether the distribution of valence bonds
with length $(x,y)$ is isotropic, as suggested by the field theory
predictions. This can be carried out by simulating different
two-dimensional clusters with periodic boundary conditions. Here, we
study clusters with $N=16^2$, $N=32^2$, $N=64^2$ and $N=128^2$ and
provide an example of the distribution $P(x,y)$ of valence bonds with
length $(x,y)$ for the biggest cluster in the inset of
Fig.~\ref{fig:prVBD2d}.
Since we expect a power-law decay in
$r=\sqrt{x^2+y^2}$, we choose logarithmic polar coordinates for the
$x$-$y$ plane and a logarithmic color scheme for the probability
$P(x,y)$. In the figure we see that the distribution seems indeed to
be rather isotropic in the center, a behavior that is perturbed for
large $r$, due to the periodic boundary conditions.

The isotropic behavior can also be observed in the main panel of
Fig.~\ref{fig:prVBD2d}, which shows the decay of the probability for
measuring a valence bond with length $r$. We observe that almost all points
lie on the power law $r^{-3}$ as predicted by the
theory Eq.~\eqref{VBdist-d-dim}.

For very long valence bonds we can see the effect of the boundary
as a deviation from the power-law, as well as through a very small
anisotropy, characterized by a small spreading fan at large $r$.
The points on the envelope of this fan can be identified
very accurately. Whereas the upper envelope contains points of valence bonds
with lengths $(0,y)$ and $(x,0)$, the bottom envelope describes valence bonds
of lengths $(x,x\pm1)$. Note, that there are no bipartite valence
bonds of lengths $(x,x)$.

Another important issue is that the raw data for sufficiently small
$r$ in Fig.~\ref{fig:prVBD2d} fall precisely on the same curve for
all system sizes, without any extra renormalization.
Considering the original definition of the VB distribution $P(r)$,
there is no obvious reason why $P(r)$ is independent of
the system size. This, however, can be understood
rather naturally from the field theory description 
discussed in Sec.~\ref{sec:sigma}.
Once the VB occupation number is mapped to a correlation
function, the latter does not
depend on the system size as long as the distance
$r$ is much smaller than the system size $L$.
In the present problem, the correlation function is not
of a standard type, but in the scaling limit
it is nonetheless mapped to the boundary correlation function
in the field theory.
Thus it should not depend on the system size,
assuming that the point-splitting scale $\epsilon$
is insensitive to $L$.
The mapping to the field theory works only for a sufficiently
large $r$.
Nevertheless, the system-size independence of the VB distribution
$P(r)$ at the shortest distance $r=1$
can be understood even without the field theory mapping.
That is, $P(1)$ is related exactly to
the ground state energy $E_0$ of the underlying
Hamiltonian Eq.~\eqref{eq:H} as
\begin{equation}
\frac{E_0}{NJ} = -\frac{1}{4}\left[P(1)+d\right].
\end{equation}
Since the ground state energy per site, $E_0/N$,
is an intensive quantity, so is $P(1)$.
It is also natural to expect that $P(r)$ for an arbitrary
distance $r$ is also intensive, which is indeed seen
in the numerical results.

The sum rule Eq.~\eqref{eq.Pr_sumrule} requires a constraint on
the asymptotic power-law behavior $P(r) \sim \calC r^{-\alpha}$
at large distances, which follows from the field theory.
Let us assume that the power law is valid for $r > a$,
where $a$ is the ultraviolet ``cutoff'' scale (usually of the order
of the lattice constant) above which the field theory applies.
Furthermore, here we assume that the proportionality constant
$\calC$ is independent of the system size (length) $L$.
We first use the decomposition
\begin{equation}
 \sum_{\mathbf{r}} P(\mathbf{r}) =
\sum_{r = |\mathbf{r}| \leq a } P(\mathbf{r}) +
\sum_{r = |\mathbf{r}| > a } P(\mathbf{r}) ,
\end{equation}
where both terms in the right-hand side are non-negative by
definition.
In a finite-size system, $r$ is actually cut off
by $L$. Thus the second term, which is the large-distance
part, is estimated as
\begin{align}
\sum_{\frac{a}{L} <  x = | \mathbf{x} | < 1} \frac{\calC}{x^\alpha}
& \sim \frac{\calC}{L^{\alpha-d}}
\int_{\frac{a}{L} < x < 1}  \frac{1}{x^\alpha} \; d^d \mathbf{x} 
\notag \\
& \sim  \frac{2 \pi^{(d+1)/2}}{\Gamma(\frac{d+1}{2})}
\frac{1}{\alpha-d}  \frac{\calC}{a^{\alpha-d}},
\label{eq.ldpart}
\end{align}
where $\mathbf{x} \equiv \mathbf{r}/L$.
Only the leading singular part in $L \rightarrow \infty$
is retained in the integral
in the last manipulation, which is valid only if $\alpha > d$.
This shows that the total contribution from the
large-distance part of $P(\mathbf{r})$ is finite even in the
thermodynamic limit $L \rightarrow \infty$ and thus the
sum rule~Eq. \eqref{eq.Pr_sumrule} can be satisfied.
If $\alpha \leq d$, on the other hand,
the contribution from the
large-distance part is divergent and consequently the
sum rule will be violated.
Therefore, the sum rule~Eq. \eqref{eq.Pr_sumrule} requires that
the asymptotic large-distance power law should have
either $\alpha > d$ or the proportionality constant $\calC$
depending on the system size $L$.
Our finding that $\alpha = d+1$ and $\calC$ is system-size
independent is consistent with this constraint.

\begin{figure}[ht]
\includegraphics[width=0.95\columnwidth]{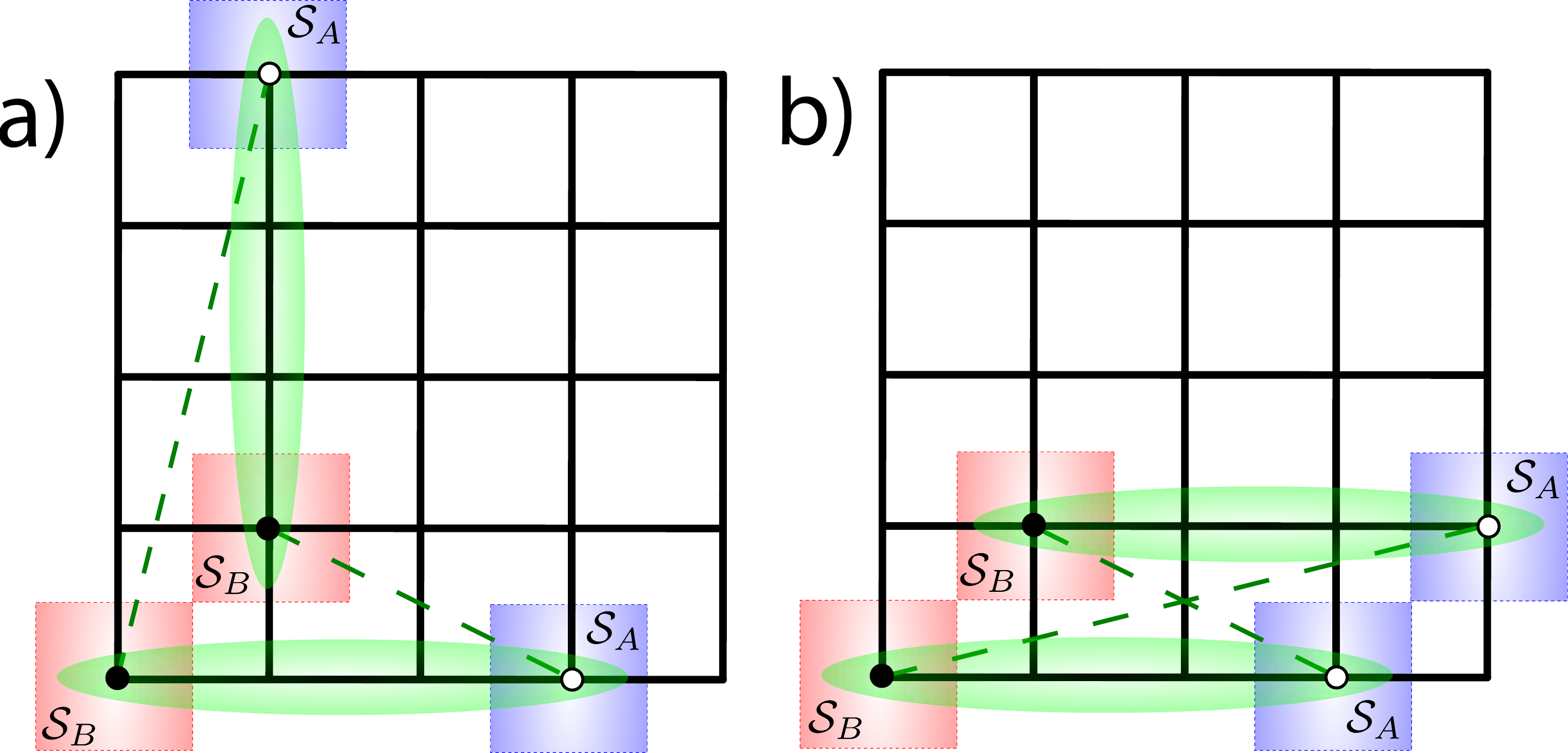}
\caption{(Color online) a) Choice of the swapped regions ${\cal S}_A$
  and ${\cal S}_B$ in order to measure the correlation between two
  valence bonds (shaded ellipses) of lengths $r$, drawing an angle of
  $\alpha=\pi/2$. Notice that Eq. \eqref{eq:gencorrelations} yields
  one when two valence bonds connect ${\cal S}_A$ and ${\cal S}_B$ and
  vanishes otherwise. This is true for the illustrated VB configuration, as well as for the one that is sketched by the dashed
  lines. b) The same configuration with two valence bonds drawing an angle of
  $\alpha=0$.}
\label{fig:dimers2D}
\end{figure}

Second, we want to study VB-VB correlations in the ground
state of the Heisenberg antiferromagnet. In order to see whether
correlations are present in two dimensions we choose the two cases
that are depicted in Fig.~\ref{fig:dimers2D}, since we intuitively
expect correlations to be most pronounced when two valence bonds are very
close to each other. We refer to these two cases as those
where two valence bonds of equal length draw an angle of a) $\alpha=\pi/2$ and
b) $\alpha=0$. We shall here emphasize again that we cannot directly measure the
simultaneous occupation number of two VBs for specified pairs
of the sites, but only the sum of occupation numbers corresponding to
two possible VB patterns.
In the present case, these patterns are drawn as shaded ellipses and dashed
lines in Fig.~\ref{fig:dimers2D}, respectively. In the case a)
$\alpha=\pi/2$, for example, we have two valence bonds of lengths $(r,0)$ and
$(0,r)$ and two other valence bonds of lengths $(r-1,1)$ and $(1,r+1)$. In the case b) we encounter the same valence bond lengths, but at a different angle $\alpha=0$.

Let $C_{\alpha}(r)$ with $\alpha=0, \pi/2$
represent the simultaneous VB occupation numbers
in the two cases a) and b), respectively.
If there is no correlation between the two VBs,
the normalized simultaneous VB occupation number on the pair
should be given as
\begin{align}
& C_{\alpha=\pi/2}(r) = C_{\alpha=0}(r)
\notag \\
& = P(r)^2 + P(\sqrt{r^2+2r+2})P(\sqrt{r^2-2r+2}),
\label{eq:Corrs2D}
\end{align}
where we used the isotropy of the single VB occupation number 
$P({\bf r})=P(r)$
on a given pair of sites separated by ${\bf r} = (x,y)$.
In fact, there are no other patterns that
are expected to have exactly the same decomposition as in
Eq.~\eqref{eq:Corrs2D}.

In order to verify Eq.~\eqref{eq:Corrs2D}, we have to find the proper
constant of proportionality. Here it is important, that we used
translational and rotational invariance in order to measure $P(x,y)$,
$C_{\alpha=0}(r)$ and $C_{\alpha=\pi/2}(r)$, accounting for a factor
of $N$ in all cases. For this reason it is obvious that the
appropriate constant of proportionality must be $1/N$. Therefore, we
have to multiply the correlation data by the system size, when comparing different system sizes $L$, since $P(x,y)$ does not depend on $L$. Furthermore, patterns of $C_{\alpha=0}(r)$ appear only half
as much as those for $C_{\alpha=\pi/2}(r)$, due to rotational
symmetry.

\begin{figure}[ht]
\includegraphics[width=0.99\columnwidth]{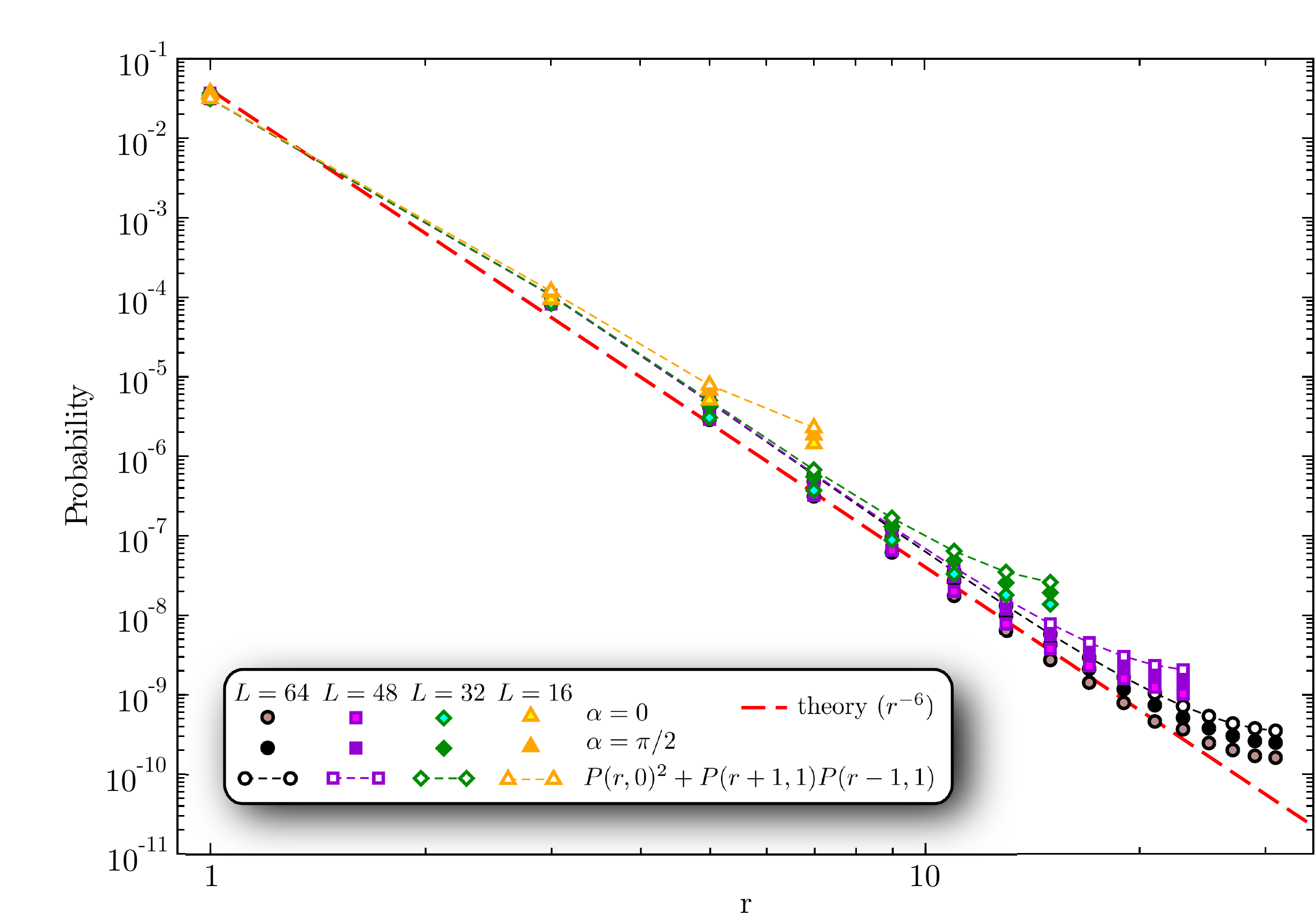}
\caption{(Color online) Simultaneous occupation number of two VBs
  on bonds that draw an angle $\alpha$ between them, as illustrated in
  Fig.~\ref{fig:dimers2D}.
  For comparison, we also show the sum of the products
  of corresponding single VB occupation numbers
  as in the right-hand side of Eq.~\eqref{eq:Corrs2D}.
  Data were obtained for
  clusters of $N=L^2$ sites with a projection length of $m/N=16$
  (convergence checked) and error bars are smaller than the point
  sizes.
  The good agreement among the three different sets of data
  confirms
  the decomposition as in Eq.~\eqref{eq:Corrs2D}, which
  indicates the asymptotic absence of the correlation between
  two VBs.
  This also implies the agreement with
  the field theory prediction $\propto r^{-6}$, as shown. 
}
\label{fig:VBC2D}
\end{figure}

The properly normalized data are plotted in Fig.~\ref{fig:VBC2D}, and show essentially the expected
$\sim r^{-6}$ behavior. This power law is also obtained by the simple
product of the pure valence bond occupations, which indicates the
absence of correlations between VBs
in two dimensions. Interestingly, the
very first data point for $r=1$ seems to be
slightly above the predicted power law $\sim r^{-6}$.
In fact, this point is the only one where the correlation
data are slightly above the sum of the products of the occupation numbers
for the individual VBs. In the entire rest of the range, this is not the
case. This implies the existence of a correlation among VBs at
very short distances $r \sim 1$.
Such an effect seems to be in agreement with some recent
results, which conclude that including correlations between valence bonds
at very short distances can indeed improve the energy for variational
calculations~\cite{Lin12}.
In any case, this does not contradict the field theory analysis,
which only applies to the asymptotic behavior at large distances.
In fact, the present numerical analysis demonstrates that
the field theory works very well above the rather short
length scale $r > a \sim \sqrt{2}$.

For larger $r$ comparable to the system size $L$,
we see the effects of the periodic boundary
conditions, and the data do not follow the field theoretical prediction
for the infinite system any more.
We furthermore see a more pronounced splitting between the
different data for $C_{\alpha=\pi/2}(r)$, $C_{\alpha=0}(r)$ and the
valence bond occupation product. However, these differences are very
small (of order $10^{-10}$) compared to the correlations
between short valence bonds.

\subsection{$d=3$ QMC results}\label{sec:QMC3d}

\begin{figure}[ht]
\includegraphics[width=0.99\columnwidth]{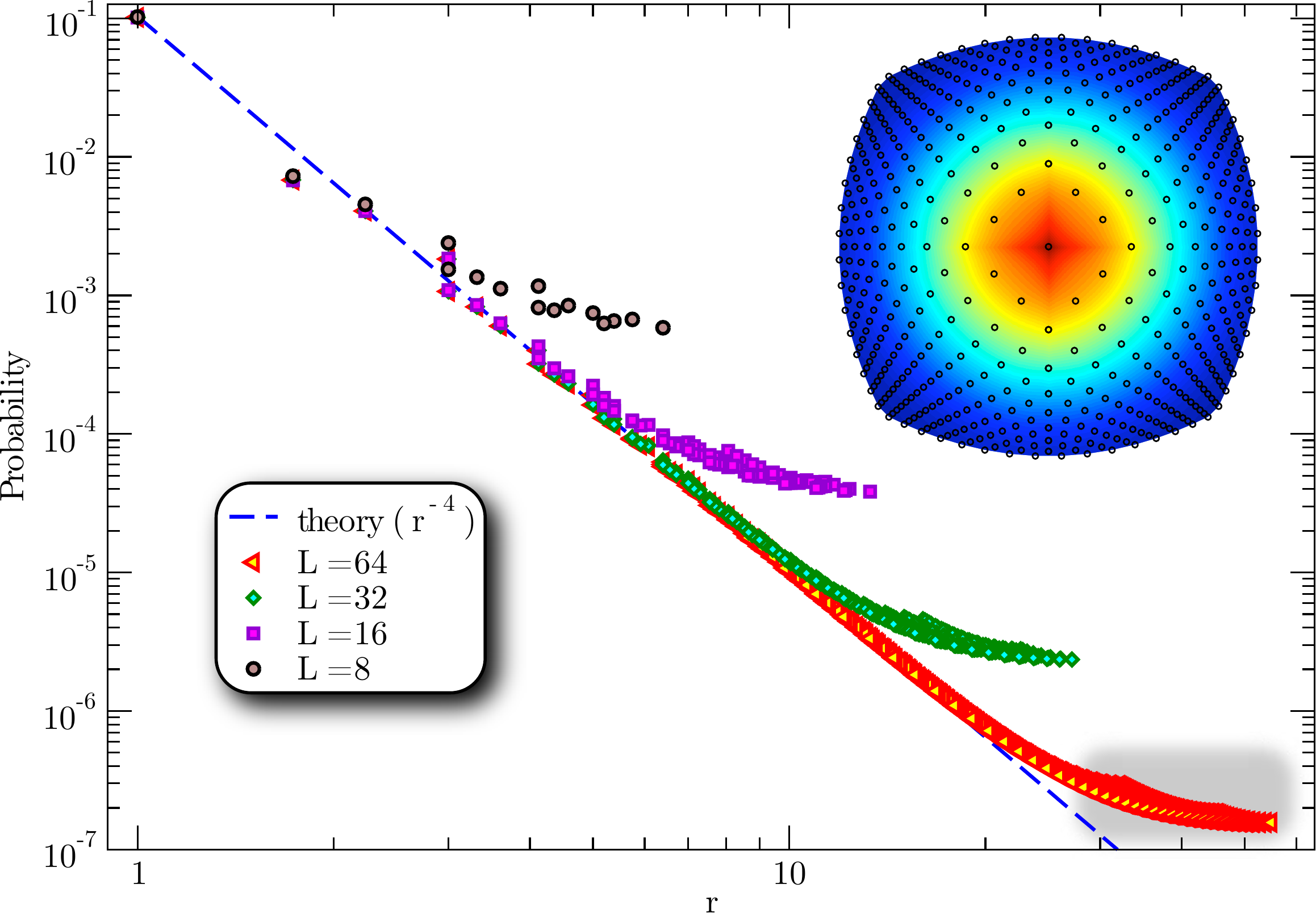}
\caption{(Color online) Length distribution $P(r)$ of valence bonds of
  length $r=\sqrt{x^2+y^2+z^2}$ (in all orientations) on different
  clusters with $N=L^3$ sites. The dashed blue line is a guide to the
  eyes, showing the expected $r^{-4}$ behavior from field
  theory. Notice that qualitatively the only difference from the case
  $d=2$ (Fig. \ref{fig:prVBD2d}) is the exponent in the observed
  power law, due to the higher dimension. Data have been produced for
  a projection length of $m/N=4$ for the two smallest lattices and  $m/N=8$ for $L=32$ (convergence was checked) and error bars
  are smaller than the point sizes.  For $L=64$ we used a relatively small projection length of $m/N=2$, resulting in non-converged data for $r\gtrsim25$ (highlighted in gray). Inset: Valence bond distribution
  $\log P(\log(r)\cos\theta,\log(r)\sin\theta)$ in logarithmic polar
  coordinates with logarithmic color scheme for $N=32^3$ in the region
  $[-L/2,L/2]^2$ within the y-z plane (at $x=0$). The data grid is
  illustrated by small black circles, whereas the colored mesh
  represents interpolated data. Similarly as for $d=2$, for short
  valence bonds the distribution appears to be rather isotropic, in
  agreement with the main panel, whereas the boundary effects of
  the finite-size sample can be observed for very long valence bonds.}
\label{fig:prVBD3d}
\end{figure}

In order to check that the same field theory also applies to higher
dimensions, we here present QMC results for the case
$d=3$. Because simulations become more and more demanding as we increase
the dimension, we restrict ourselves to the distribution of valence
bonds $P(r)$, which is depicted in Fig.~\ref{fig:prVBD3d} for
samples of up to $N=64^3$ sites. Furthermore, for the biggest cluster we could not achieve convergence of the data for $r\gtrsim25$ (fortunately, this is a region where boundary effects start to play a role and where deviations from field theory predictions are therefore expected).

As expected, we observe the same behavior as in two dimensions, but with the exponent in the
power law now being equal to $4$. We furthermore observe that the
distribution is still isotropic in three dimensions, as can be seen in
the inset of Fig.~\ref{fig:prVBD3d}, where a two-dimensional cut
through the data is shown.

\section{Asymptotic behavior of VB occupation numbers
and their correlations in $d = 1$}

\subsection{Boundary field theory approach in $d=1$: Bosonization}
\label{sec:bos}

The physics of the $S=1/2$ Heisenberg antiferromagnet in a $d=1$
chain is quite different from that of the same model in the $d \geq 2$
hypercubic lattice. In $d=1$, quantum fluctuations are so strong
that the long-range antiferromagnetic order is absent
even in the ground state. The ground state of the $S=1/2$ antiferromagnetic chain
is critical: Various correlation functions decay algebraically.
The universal behaviors in the low-energy, long-distance
regime are described~\cite{Affleck-LesHouches88}
as a Tomonaga-Luttinger liquid (TLL).
This is a free-boson field theory defined by the Lagrangian
density
\begin{equation}
\calL = \frac{1}{2} (\partial_{\mu} \phi)^2 .
\label{TLLLag}
\end{equation}
The field $\phi$ obeys the equation of motion, which is
nothing but the wave equation in one spatial dimension.
As a consequence, $\phi$ can be decomposed as
\begin{equation}
\phi(x,t) = \phi_R(x-t) + \phi_L(x+t),
\label{phiRL}
\end{equation}
where $\phi_{R,L}$ are right-moving and left-moving components.
We also introduce the dual field $\tilde{\phi}$, which is defined by
\begin{equation}
 \tilde{\phi} = \phi_R - \phi_L .
\end{equation}

The spin operators are represented in terms of the boson field $\phi$ as
\begin{align}
S^z_j & \sim \frac{1}{2\pi R}\frac{\partial \phi}{\partial x}
+ \const (-1)^j \cos{\frac{\phi}{R}} ,
\label{Sz_in_phi}
\\
S^{\pm}_j & \sim \const
 e^{ \pm 2 \pi R i \tilde{\phi}}
 \cos{\frac{\phi}{R}} + \const (-1)^j e^{ \pm 2 \pi R i \tilde{\phi}},
\label{Spm_in_phi}
\end{align}
where $R$ is a parameter of the theory called 
the compactification radius.
For the isotropic Heisenberg antiferromagnetic chain,
\begin{equation}
 R = \frac{1}{\sqrt{2\pi}} .
\end{equation}

Despite the significant difference in physics
between $d=1$ and $d \geq 2$, it is possible to
construct a boundary field theory formulation
of VB distributions in $d=1$ as well.
First, let us identify the conformally invariant boundary
condition, which would be the infrared fixed point for
the given boundary state $| \mbox{\Neel} \rangle$.
In the reference {\Neel} state, the staggered component of
$S^z$ is fixed to a constant.
Thus it is natural to expect that the Dirichlet boundary
condition $\phi=0$ is the appropriate boundary condition.
The Dirichlet boundary conditon is in fact conformally invariant.
Assuming that this is the case,
the valence bond occupation number is just given by
the correlation function $\langle S^+ S^- \rangle$
along the boundary, where 
the Dirichlet boundary condition $\phi = 0$ is imposed.
This problem is reduced to the correlation function
of the vertex operators in the presence of the
Dirichlet boundary condition:
$\langle e^{2 \pi R i \tilde{\phi}} e^{2 \pi R i \tilde{\phi}}\rangle_D$.

Such boundary correlation functions have been studied
in great detail in boundary conformal field theory.
In the present case with the Dirichlet boundary condition,
the calculation can be quite easily done as follows.
Let us assume the system is defined on the half plane $\tau < 0$
and the boundary is at $\tau =0$, along the $x$ axis.

Because of Eq.~\eqref{phiRL},
imposing the Dirichlet boundary condition $\phi = 0$
at the boundary $\tau=0$ is equivalent to extending  
the chiral field $\phi_R$ to the other side of the boundary $\tau > 0$,
where no field was originally defined, by the following relation:
\begin{equation}
 \phi_R(x, \tau) = - \phi_L( x, - \tau). 
\end{equation}
This means that the dual field at the boundary $\tau=0$
can be written entirely in terms of the chiral field $\phi_R$.
Moreover, the correlation function of the chiral field
can be evaluated in the infinite plane without boundary.
Combining this observation with
Eqs.~\eqref{eq.vb_as_bcf} and \eqref{Spm_in_phi},
the large distance asymptotic behavior of the valence
bond occupation number is given as
\begin{equation}
\bar{n}_{jk} \propto
\langle e^{i 4\pi  R \phi_R} e^{-i 4\pi R \phi_R} \rangle,
\end{equation}
where $\langle \cdot \rangle$ now means the expectation value in
the infinite plane without a boundary.
The result can be readily obtained as
\begin{align}
\langle e^{i 4\pi  R \phi_R} e^{-i 4\pi R \phi_R} \rangle
&=
\const \frac{1}{r^{4 \pi R^2}}
\\
&= \const \frac{1}{r^2},
\label{eq:distribution1D}
\end{align}
where $r$ is the distance between the two sites $j$ and $k$,
and we have set $R=1/\sqrt{2\pi}$ for the isotropic
Heisenberg chain in the second line.
Despite the difference in the formulation,
the final result for the single VB occupation number
in $d$ dimensions can be summarized as $1/r^{d+1}$,
whether $d=1$ or $d \geq 2$.

On the other hand, correlations among VBs reveal an
interesting difference between $d=1$ and $d\geq 2$.
Although the effective field theory (TLL) is also a free-boson
field theory, (the staggered part of) the spin operator is expressed
by a vertex operator (exponential in the boson field), and not by
the boson field itself.
Thus, while Wick's theorem certainly applies to the TLL,
it does not mean the absence of correlations in multipoint
correlation functions of {\em spins}, and consequently, among
the VB occupations.

In fact, the simultaneous occupation number of
two VBs [Eq.~\eqref{eq.2VBocc}] can be evaluated exactly
within the TLL theory formulation Eq.~\eqref{TLLLag}.
Following the same logic as in the single VB occupation number,
we find
\begin{align}
& \bar{n}_{\{x_1,x_2 \},\{ x'_1,x'_2 \}}
\notag \\
& \propto
\langle
e^{4\pi R i \phi_R(x_1) }
e^{4\pi R i \phi_R(x_2) }
e^{- 4\pi R i \phi_R(x'_1) }
e^{- 4\pi R i \phi_R(x'_2) }
\rangle
\notag \\
& \propto
\left(
\frac{r_A r_B}{
r_{11}r_{22}r_{12}r_{21}}
\right)^2,
\end{align}
where $x_1,x_2 \in A$, $x'_1,x'_2 \in B$, and
\begin{align}
r_{mn} &= | x_m - x'_n |, \\
r_A & =  |x_1 - x_2 |, \\
r_B & =  |x'_1 - x'_2 | .
\label{eq:correlation1D}
\end{align}
Again we have assumed $R=1/\sqrt{2\pi}$ for the SU(2) symmetric
Heisenberg antiferromagnetic chain.

When two bonds are separated by a large distance (relative to the
bond lengths), namely, in the limit 
\begin{equation}
r_A \sim r_B \sim r_{12} \sim r_{21} \gg r_{11} \sim r_{22}, 
\end{equation}
we find
\begin{equation}
\bar{n}_{\{x_1,x_2 \},\{ x'_1,x'_2 \}}
\sim 
\left(
\frac{1}{r_{11}r_{22}}
\right)^2 .
\end{equation}
This implies that
\begin{equation}
\bar{n}_{\{x_1,x_2 \},\{ x'_1,x'_2 \}}
\sim 
\overline{n_{(x_1,x_2)}} . 
\overline{n_{(x'_1,x'_2)}}   .
\end{equation}
Namely, the correlation asymptotically
vanishes, as the separation between two VBs is taken to infinity.
This is consistent with our intuitive expectation.

On the other hand, for general separation, our result exhibits
a nontrivial correlation between VB occupation.
A major feature of the correlation is that,
because of the factor $r_A r_B$ in the
numerator, the simultaneous occupation amplitude
vanishes as $x_1 \rightarrow x_2$ or $x_1' \rightarrow x_2'$.

\subsection{$d=1$ QMC results}\label{sec:QMC1d}
\begin{figure}[htbp]
\begin{center}
\includegraphics[width=0.99\columnwidth]{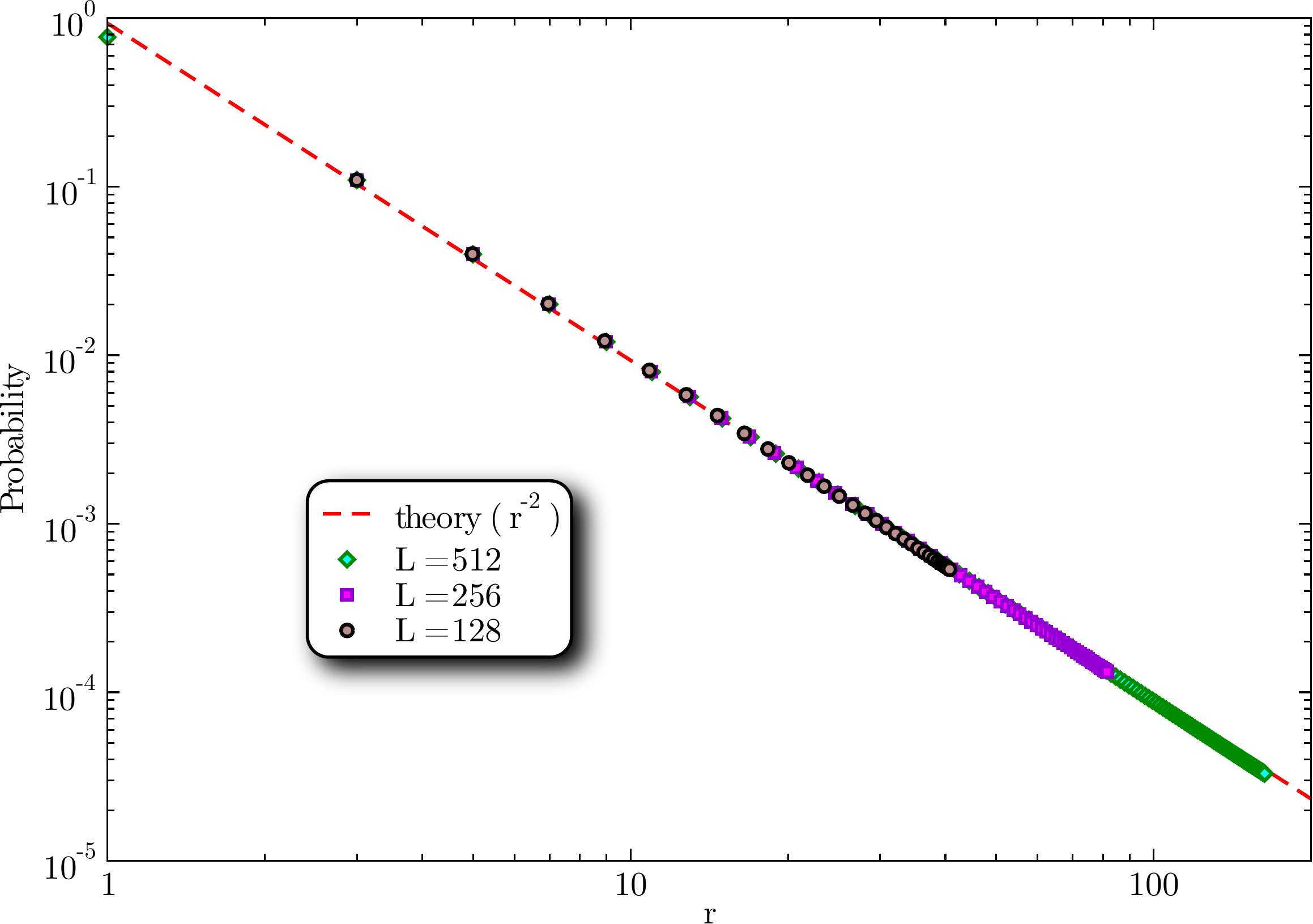}
\caption{(Color online) Length distribution $P(r)$ of valence bonds as a function of $r$ for 
  one-dimensional lattices with $L$ sites as obtained from QMC calculations, versus field theory
  predictions. All chains were simulated for a projection length of
  $m/L=64$ and the convergence was checked (error bars are smaller
  than the point sizes). Note that we applied the conformal
  transformation $r=L/\pi\sin(\pi x/L)$, where $x$ is the measured valence bond
  length.}
\label{fig:VBD}
\end{center}
\end{figure}

Let us now compare the theoretical predictions above
with the numerical results obtained by QMC simulations.
As we have seen earlier, QMC simulations are restricted
to finite sizes, and the VB distribution functions
inevitably deviate from the theoretical predictions
which were made for an infinite system.
Fortunately, however, our theory for $d=1$ is based on
the TLL, which is a conformal field theory
in $1+1$ dimensions. In this class of field theory,
the finite-size effect is conveniently described
by a conformal mapping~\cite{YellowPages}.
The rule of thumb is that, in order to obtain
a correlation function in a finite system (ring)
of length $L$ with the periodic boundary condition,
a distance $r$ appearing
in the corresponding correlation function
in the infinite system should be replaced by
the arc distance $r = (L/\pi) \sin{(\pi x/L)}$
on the finite ring.
In the one-dimensional case we also first check the length distribution
$P(r)=\bar{n}_{k,l}$ of the valence bonds,
as plotted in Fig.~\ref{fig:VBD}. The simulations are
carried out on the one-dimensional Heisenberg chain with $L=128$, $256$ and
$512$ sites and periodic boundary conditions. We find a power-law decay
$P(r) \sim \calC r^{-2}$, which agrees very well with
the field theory prediction Eq.~\eqref{eq:distribution1D}
with the conformal mapping.

Here again, all the data in Fig.~\ref{fig:VBD} for different
system sizes appear to collapse on a single curve,
without any rescaling.
That is, the normalization constant $\calC$ is
independent of the system size.
As we have discussed in Sec.~\ref{sec:QMC2d}, this can
be naturally understood from the mapping to the 
boundary correlation function and 
the intensiveness of the ground state energy per site.
In fact, we can make the analysis in Eq. \eqref{eq.Pr_sumrule}
more precise, by using the correlation function in a
finite system of length $L$ obtained by the conformal mapping, as
\begin{align}
\calC \sum_{\frac{a}{L} <  x = | \mathbf{x} | < 1}
\left( \frac{1}{L \sin{(\pi x )}} \right)^\alpha
& \sim \frac{\calC}{L^{\alpha -1}} \int_{\frac{a}{L} < x < 1} 
\left(  \frac{1}{\sin{\pi x}} \right)^\alpha \; dx
\notag \\
& \sim
\frac{2}{\alpha-1}  \frac{\calC}{a^{\alpha-d}}.
\end{align}
The leading contribution in the limit of $L\rightarrow \infty$,
for $\alpha > 1$,
however remains the same as in~Eq. \eqref{eq.Pr_sumrule}
with $d=1$. For $\alpha < 1$, the large-distance contribution diverges,
as in the case of $\alpha < d$ for general dimension $d$.
Our finding that $\calC$ is independent of the system
size and that $\alpha = d+1$ applies also to $d=1$,
again in consistency with the sum rule~Eq. \eqref{eq.Pr_sumrule}.




Only at the shortest distance $r=1$, there is a visible
difference between the numerical result and the
field theory prediction.
As in $d \geq 2$, the disagreement between the data and
the field theory prediction is not surprising,
as the field theory describes
only the long-distance asymptotic behaviors.
The data imply that the ultraviolet cutoff scale $a$,
above which the field theory is valid,
is small ($\sim 1$) also in $d=1$.

\begin{figure}[ht]
\includegraphics[width=0.95\columnwidth]{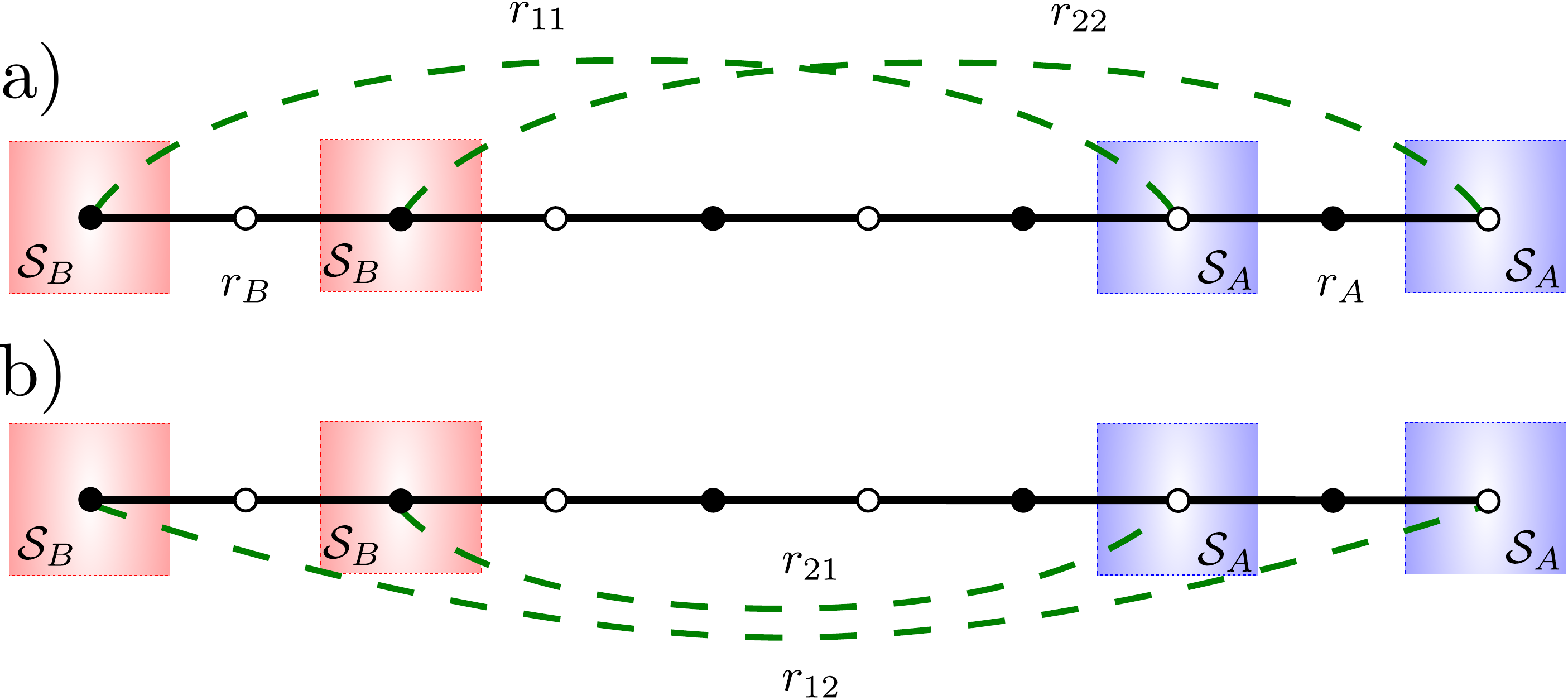}
\caption{(Color online) a) Choice of the disconnected swapped regions ${\cal S}_A$ (separated by $r_A=2$) and ${\cal S}_B$ (separated by $r_B=2$) with the corresponding valence bonds of lengths $r_{11}$ and $r_{22}$. b) There is also another possible configuration of bipartite valence bonds with lengths $r_{12}$ and $r_{21}$.}
\label{fig:dimers1D}
\end{figure}

\begin{figure}[htbp]
\begin{center}
\includegraphics[width=0.99\columnwidth]{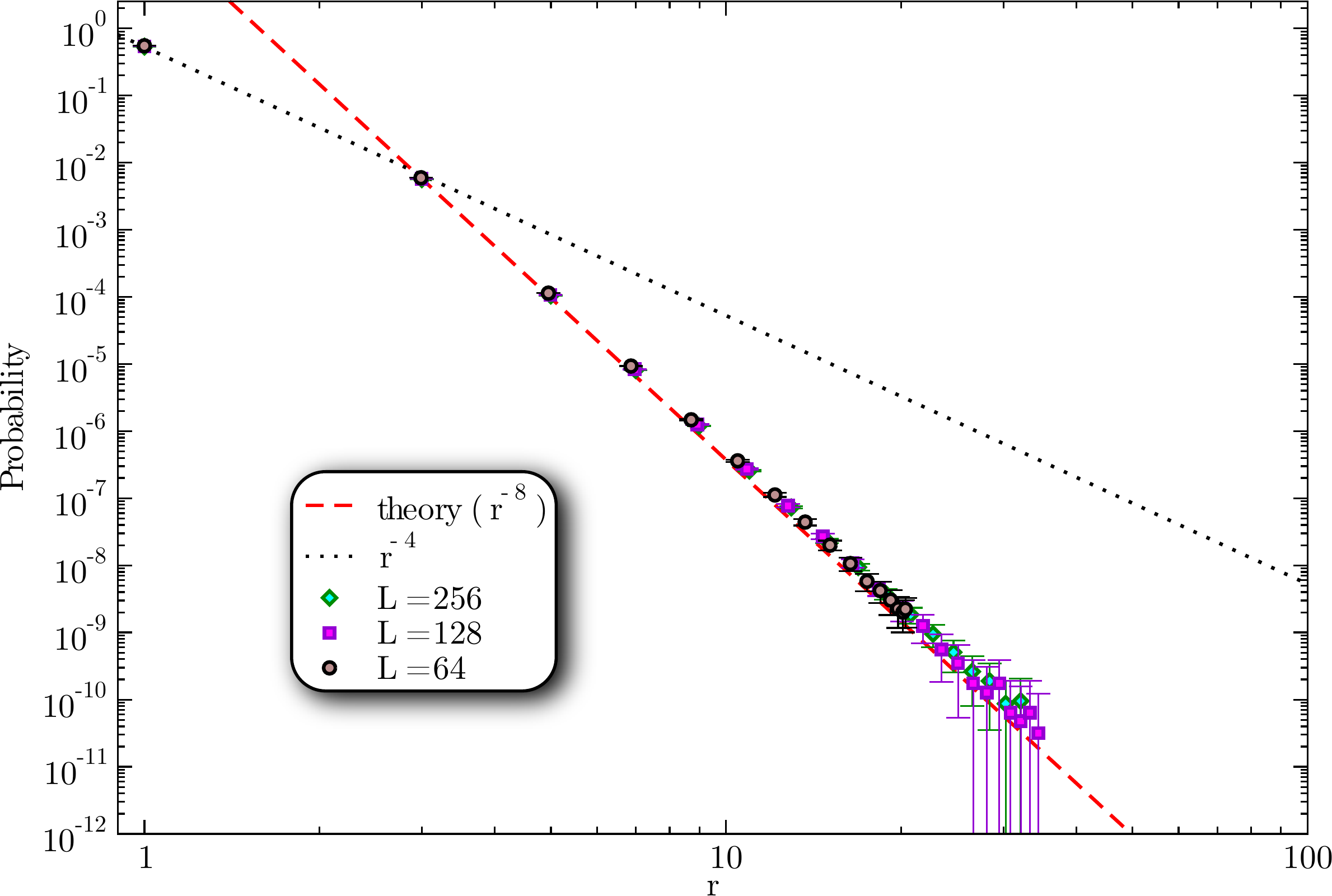}
\caption{
(Color online) Correlations between valence bonds (in the configuration of Fig.~\ref{fig:dimers1D}) in the ground state of Eq.~\eqref{eq:H}  on $1D$ chains of different sizes $L$. The measured correlations multiplied by system size (data points)  clearly agree with the field theory predictions of a power-law decay $r^{-8}$ (red dashed curve). All data were obtained with projection lengths of $m/L=64$ (convergence was checked) and the error bars are smaller than the point sizes if omitted. Notice, that we applied the conformal transformation $r=L/\pi\sin(\pi x/L)$, with $x$ being the measured distances.}
\label{fig:VBC1D}
\end{center}
\end{figure}

We now study the correlations between VBs on the Heisenberg chain,
where we choose again the case of two equally long close valence bonds, as
illustrated in Fig.~\ref{fig:dimers1D}. Again, there are two
possible VB patterns and we choose $x=r_{11}$ as the varying parameter.
The results are plotted in
Fig.~\ref{fig:VBC1D}. Notice first of all, that
Eq.~\eqref{eq:correlation1D} uses conformal coordinates, such that we
expect a power-law decay as $\sim r^{-8} \propto \sin^{-8}(\pi x/L)$ for
sufficiently long valence bonds. This implies, however, a rather drastic
decrease of probabilities, resulting in relatively rare events. This
is indeed observed in the simulations, where probabilities of
$10^{-11}$ with pronounced error bars are encountered. Nevertheless,
we can confirm the predicted $\sim r^{-8}$ behavior and clearly exclude
a possible $\sim r^{-4}$ power-law decay obtained from a simple product
of valence bond occupations $C_{\rm No \quad \!\!\!\!\! corr}(x) =
P(x)^2 + P(x-2)\cdot P(x+2)$. This is a strong indication, that
correlations are indeed present
even at large distances, in one dimension.

\begin{figure}[htbp]
\begin{center}
\includegraphics[width=0.99\columnwidth]{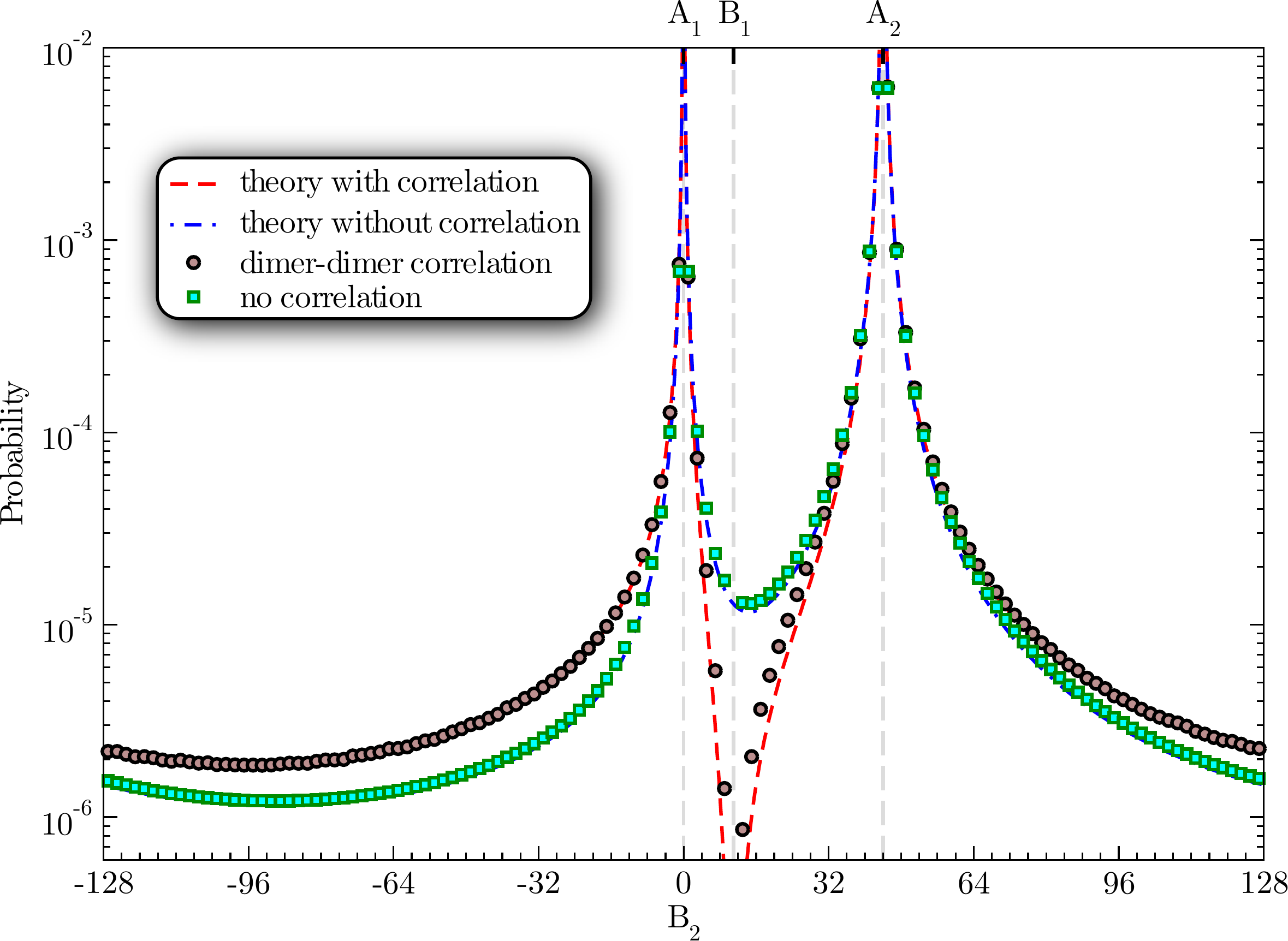}
\caption{(Color online) VB-VB correlations on a chain with $L=256$ sites for the two (undistinguishable) configurations $\fullket{(A_1,B_1)(A_2,B_2)}$ and $\fullket{(A_1,B_2)(A_2,B_1)}$, with $A_1=0$, $A_2=44$, $B_1=11$ and $B_2=x$ varying. We show the product of valence bond occupations (green squares), that fit to a theory without correlation as comparison. However, the measured correlations (multiplied by the system sizes, shown as black circles) rather correspond to a theory with correlations. Notice, how the two theories agree asymptotically around $A_1$ and $A_2$ and disagree substantially at $B_1$. The projection length is $m/L=64$ (convergence was checked) and error bars are smaller than point sizes.}
\label{fig:VBCpumps}
\end{center}
\end{figure}

In order to further examine this behavior, we now study a case where three points of this four-point correlation function are fixed and one point is variable, as shown in Fig.~\ref{fig:VBCpumps}. We plot the VB-VB correlations, as well as the product of the valence bond occupations, which both reproduce the theoretical predictions. In agreement with our naive intuition, both theories show the same behavior if we consider short valence bonds, separated by a rather large distance. This can be seen in the figure when $x\approx A_1$ or $x\approx A_2$ and means that no further correlations are present in such a case. However, the two curves differ considerably for $x\approx B_1$, i.e., when the two valence bonds start affecting each other. This confirms that in such a case correlation effects are clearly important in one dimension.

\section{Discussion and conclusion}\label{sec:discussion}

Through a combination of analytical calculations and quantum Monte Carlo simulations, we have investigated the behavior of valence bond probability distributions and correlations in ground states of the antiferromagnetic Heisenberg model on chain, square and cubic lattices. 

Our analytic results provide a clear ground for the probability distribution $P({\bf r})$ for a valence bond to join two sites (existing on different sublattices) separated by $\mathbf{r}$ to scale as $|\mathbf{r}|^{-(d+1)}$ where $d$ is the lattice dimension. This result, which was observed in previous numerical simulations and justified by a mean-field ansatz, is now understood within a more trustworthy analytical framework: bosonisation in 1D, and the non-linear $\sigma$ model in $d>1$. This will be useful for variational calculations, or for constructing guiding wave functions in Monte Carlo simulations, which aim at targeting an antiferromagnetic state on a bipartite lattice.

Moreover, our results provide a formal justification for the
factorization ansatz Eq.~\eqref{eq:LDA} introduced by Liang,
Dou\c{c}ot and Anderson, as long as a non-linear $\sigma$ model
description is valid. Indeed, our results, corroborated by large-scale
QMC calculations in the 2D case, clearly indicate the absence of
correlations between valence bonds at long distances. For the
quasi-long-range ordered ground state of the Heisenberg chain, the
situation is more subtle as correlations are present between
valence bonds, as clearly observed in Fig.~\ref{fig:VBCpumps}. This may be understood as a signature of the
strong (power-law-decaying) dimer correlations which are known
\cite{Giamarchi} to be present in the Heisenberg AF chain.

We would also like to point out that the existence of long-range
antiferromagnetic order does not necessarily imply a distribution of
valence bonds with a distribution law $P({\bf r}) \propto |{\bf
  r}|^{-(d+1)}$ with no correlations. For instance, the wave functions
composed of the equal-weight linear combinations of nearest-neighbor
valence bonds on the simple cubic or diamond lattices have been shown
to sustain AF long-range order~\cite{Albuquerque12} (see also Ref.~\onlinecite{Xu13}). However, their
description is most certainly beyond the non-linear $\sigma$ model
approach, which probably cannot describe the dipolar dimer
correlations observed in these wave functions~\cite{Albuquerque12}.

Finally, we suggest several further investigations as possible
extensions of our work. For one-dimensional systems, the
bosonization analysis of Sec.~\ref{sec:bos} carries over for the
critical phase of the $XXZ$ anisotropic spin chain. While the SU(2)
symmetry is lost in this more general case, analytical predictions for
mixed expectation values of the type $\langle \mbox{N\'eel}|
S^+_iS^-_jS^+_kS^-_l | \psistate \rangle$ [such as in
Eq.~\eqref{eq:gencorrelations}] can be made and tested through, for
instance, DMRG or QMC calculations. In general, the power-law exponent
will depend on the anisotropy parameter through the compactification
radius $R$. Another interesting problem would be
to calculate analytically logarithmic corrections to the power-law
decays of valence bond occupations and correlations in one dimension,
which most certainly exist in the SU($2$) case. We note, however, that
these are presumably small, hard to detect, effects as the QMC
results are already very well described by the pure power-law decays.

In dimensions larger than 1, the non-linear $\sigma$ model approach may
also be used to describe the behavior of valence bond distributions
and correlations at a quantum critical point between the
antiferromagnet and a paramagnet, such as for a bilayer Heisenberg
model~\cite{SandvikScalapino,SandvikChubukov}.
This could be useful in explaining and improving variational approaches based on valence bonds
that aim to describe this quantum phase transition (such as, {\it e.g.},
Ref.~\onlinecite{Liao}).
More generally, the current formalism based on the
overlap Eq.~\eqref{eq:gencorrelations} could be used
with any effective field theory which describes
a particular quantum phase or a quantum critical point
of quantum antiferromagnets on bipartite lattices.

Finally, it would be very interesting to see whether an approach
similar to the one developed here could be applied to describe the
``spinon detection'' procedure recently advocated by Tang and
Sandvik~\cite{Tang}, which is also based on a valence bond description
of a spin system with one or two unpaired spins ({\it i.e.}, that do
not belong to a valence bond). This is a more challenging case as one
would first need to derive equations similar to Eq.~\eqref{eq:nkldef} or~\eqref{eq:gencorrelations} for this situation.

\section*{Acknowledgments}
We thank Matthieu Mambrini for very insightful discussions and participation at an early stage of this project. This work was performed using HPC resources
from GENCI-CCRT, GENCI-IDRIS (grants x2010050225, x2011050225, x2012050225) and CALMIP (grants 2011-P0677 and 2012-P0677) and is supported by
the French ANR program ANR-08-JCJC-0056-01, the
Indo-French Centre for the Promotion of Advanced Research
(IFCPAR/CEFIPRA) under Project 4504-1, and MEXT/JSPS KAKENHI Grant
Nos. 20102008 and 25400392.  M.~O. thanks Laboratoire de Physique
Th\'{e}orique, IRSAMC, Universit\'e de Toulouse and CNRS
for hospitality during his visits, during which the present
work was initiated and then completed.


\begin{thebibliography}{99}

\bibitem{BetheMerminWagner} H. Bethe, Z. Phys. {\bf 71}, 205 (1931); N.D. Mermin and H. Wagner, Phys. Rev. Lett. {\bf 17}, 1133 (1966)

\bibitem{LRO} E. Jord\~ao Neves and J. Fernando Peres, Phys. Lett. A {\bf 114}, 331 (1986); I. Affleck, T. Kennedy, E.H. Lieb and H. Tasaki, Commun. Math. Phys. {\bf 115}, 477 (1988); F.J. Dyson, E.H. Lieb and B. Simon, J. Stat. Phys. {\bf 18}, 335 (1978)

\bibitem{NumLRO} M.S. Makivi\'c and H.-Q. Ding, Phys. Rev. B {\bf 43}, 3562 (1991)

\bibitem{SandvikEvertz} A.W. Sandvik and H.-G. Evertz, Phys. Rev. B. {\bf
    82}, 024407 (2010).

\bibitem{LiebMattis} E. H. Lieb and D. C. Mattis, J. Math. Phys. {\bf 3}, 749 (1962). 

\bibitem{old} G. Rumer, E. Teller and H. Weyl, Nachr. Ges. Wiss. Goettingen, MathPhys. Kl. {\bf 499} (1932);  L. Hulth\'en, Ark. Mat. Astron. Fys. B {\bf 26A}, 1 (1938).

\bibitem{Anderson} P.W. Anderson, Mater. Res. Bull. {\bf 8}, 153 (1973); P. Fazekas and P.W. Anderson, Philos. Mag. {\bf 30}, 423 (1974).

\bibitem{TriangularLRO} R.R.P. Singh and D.A. Huse, Phys. Rev. Lett. {\bf 68}, 1766 (1992); B. Bernu {\it et al.}, Phys. Rev. B {\bf 50}, 10048 (1994); L. Capriotti, A.E. Trumper, and S. Sorella, Phys. Rev. Lett. {\bf 82}, 3899 (1999); S.R. White and A.L. Chernyshev, Phys. Rev. Lett. {\bf 99}, 127004 (2007) 

\bibitem{MoessnerSondhi}
R. Moessner and S.~L. Sondhi, Phys. Rev. Lett. {\bf 86}, 1881 (2001).

\bibitem{MisguichSerbanPasquier}
G. Misguich, D. Serban, and V. Pasquier, Phys. Rev. Lett. {\bf 89},
137202 (2002).


\bibitem{kagome} See {\it e.g.} M. Mambrini and F. Mila, Eur. Phys. J. B  {\bf 17}, 651 (2000).

\bibitem{Mambrini06} M. Mambrini, A. L\"auchli, D. Poilblanc and F. Mila, Phys. Rev. B {\bf 74}, 144422 (2006) 

\bibitem{LouSandvik} J. Lou and A.W. Sandvik, Phys. Rev. B {\bf 76}, 104432 (2007).

\bibitem{Zhang13} X. Zhang and K.S.D. Beach, Phys. Rev. B {\bf 87}, 094420 (2013).

\bibitem{old2} R. Saito, J. Phys. Soc. Jap. {\bf 59}, 482 (1990);
  H.N. Temperley and E.H. Lieb, Proc. Roy. Soc. Lond. A. {\bf 322}, 251
  (1971)

\bibitem{AndersonBook} P.W. Anderson, {\it Basic Notions of Condensed Matter Physics} (Benjamin, New York, 1984).

\bibitem{BeachSandvik} K.S.D. Beach and A.W. Sandvik, Nucl. Phys. B {\bf 750}, 142 (2006).

\bibitem{Mambrini} M. Mambrini, Phys. Rev. B {\bf 77}, 134430 (2008).

\bibitem{LDA} S. Liang, B. Dou\c{c}ot and P.W. Anderson, Phys. Rev. Lett. {\bf 61}, 365 (1988)


\bibitem{Sandvik05} A.W. Sandvik, Phys. Rev. Lett. {\bf 95}, 207203 (2005) 

\bibitem{BeachMaster} K.S.D. Beach, Phys. Rev. B {\bf 79}, 224431 (2009).

\bibitem{Beach} K.S.D. Beach, preprint arXiv:0707.0297 (2007) (unpublished).

\bibitem{Wegner} F.J. Wegner, Z. Phys. B {\bf 85}, 259 (1991).

\bibitem{note-beach} Note however that Ref.~\onlinecite{Beach} demonstrates that the amplitude product state is a correct ansatz for a long-range Heisenberg spin model, treated at the mean-field level.

\bibitem{Alet10} F. Alet, I.P. McCulloch, S. Capponi, and M. Mambrini, Phys. Rev. B {\bf 82}, 094452 (2010)

\bibitem{YellowPages}
P. Di~Francesco, P. Mathieu, and D. S\'en\'echal,
{\it Conformal Field Theory}, Springer (1997).

\bibitem{BlumenhagenPlauschinn}
R. Blumenhagen and E. Plauschinn, {\it Introduction to Conformal Field
Theory: With Applications to String Theory}, Lecture Notes in Physics
779, Springer (2009).

\bibitem{FradkinMoore}
E. Fradkin and J.~E. Moore, Phys. Rev. Lett. {\bf 97}, 050404 (2006).

\bibitem{MO-EE-BCFT}
M. Oshikawa, preprint arXiv:1007.3739 (2010) (unpublished).

\bibitem{JacobsenSaleur} J. L. Jacobsen and H. Saleur, Phys. Rev. Lett. {\bf 100}, 087205 (2008)

\bibitem{StephanMisguichPasquier}
J.-M. St\'ephan, G. Misguich, and V. Pasquier,
Phys. Rev. B {\bf 84}, 195128 (2011).


\bibitem{HuseElser} D.A. Huse and V. Elser,  Phys. Rev. Lett. {\bf 60}, 2531
  (1988) 

\bibitem{BeyondLDA} F Mezzacapo {\it et al.}, New J. Phys. {\bf 11}, 083026 (2009)
  (2009), H.J. Changlani {\it et al.}, Phys. Rev. B {\bf 80}, 245116 (2009);
  S. Al-Assam {\it et al.}, Phys. Rev. B {\bf 84}, 205108 (2011)

\bibitem{Lin12} Y.-C. Lin, Y. Tang, J. Lou and A.W. Sandvik, Phys. Rev. B {\bf 86}, 144405 (2012)

\bibitem{Zhang13} X. Zhang, J. Xu and K. S. D. Beach, preprint arXiv:1310.6030 (2013) (unpublished).

\bibitem{Marshall} W. Marshall, Proc. R. Soc. London Ser. {\bf A 232}, 48 (1955).

\bibitem{SandvikBeach} A.W. Sandvik and K.S.D. Beach, in Computer Simulation Studies in Condensed-Matter Physics XX, ed. D. P. Landau, S. P. Lewis, and H.-B. Sch\"uttler (Springer, Berlin, 2008).

\bibitem{Schwandt09} D. Schwandt, F. Alet and S. Capponi, Phys. Rev. Lett. {\bf 103}, 170501 (2009).

\bibitem{Hastings10} M. B. Hastings, I. Gonzalez, A. B. Kallin and R.G. Melko, Phys. Rev. Lett. {\bf 104}, 157201 (2010).

\bibitem{vbee} F. Alet {\it et al.}, Phys. Rev. Lett. {\bf 99}, 117204 (2007);
  R.W. Chhajlany, P. Tomczak, and A. W\'ojcik, Phys. Rev. Lett. {\bf 99},
  167204 (2007); A.B. Kallin, I. Gonz\'alez, M.B. Hastings, and R.G. Melko
Phys. Rev. Lett. {\bf 103}, 117203 (2009); H. Tran and N.E. Bonesteel,
Phys. Rev. B {\bf 84}, 144420 (2011).

\bibitem{Sutherland} B. Sutherland, Phys. Rev. B {\bf 37}, 3786 (1988).

\bibitem{note1}
Following the logic here, exchanging $S^+$ and $S^-$ 
still gives the same result in the field theory.
However, in the original formula Eq.~\eqref{eq.vb_as_bcf},
the amplitude vanishes after the exchange of the
spin operators $S^+$ and $S^-$.
This shows the presence of a subtlety in
the mapping to the field theory.
Despite this, we believe that the asymptotic large-distance limit is
correctly given by the present field-theory prescription,
for the appropriate (nonvanishing) amplitude.
This will be also supported by numerical calculations 
in Sec.~\ref{sec:QMC2d}.


\bibitem{Affleck-LesHouches88}
I. Affleck, in {\it Fields, Strings and Critical Phenomena.
Proceedings of the Les Houches Summer School 1988},
E. Br\'{e}zin and J. Zinn-Justin (eds.), North Holland (1990).

\bibitem{Giamarchi} T. Giamarchi and H. J. Schulz, Phys. Rev. B {\bf 39}, 4620 (1989).


\bibitem{Albuquerque12} A.F. Albuquerque, F. Alet, and R. Moessner, Phys. Rev. Lett. {\bf 109}, 147204 (2012)

\bibitem{Xu13} J. Xu and K. S. D. Beach, preprint arXiv:1311.0004 (2013) (unpublished).

\bibitem{SandvikScalapino} A.W. Sandvik and D. J. Scalapino, Phys. Rev. Lett. {\bf 72}, 2777 (1994) 

\bibitem{SandvikChubukov}  A.W. Sandvik, A.V. Chubukov and S. Sachdev, Phys. Rev. B {\bf 51}, 16483 (1995) 

\bibitem{Liao} H. Liao and T. Li, J. Phys.: Condens. Matter {\bf 23}, 475602 (2011).

\bibitem{Tang} Y. Tang and A.W. Sandvik, Phys. Rev. Lett. {\bf 107}, 157201 (2011); Phys. Rev. Lett. {\bf 110}, 217213 (2013).



\end{thebibliography}
\end{document}